\documentclass[12pt]{iopart}

\expandafter\let\csname equation*\endcsname\relax
\expandafter\let\csname endequation*\endcsname\relax 

\usepackage[english]{babel}
\usepackage{amsmath}
\usepackage{amssymb}
\usepackage{graphicx}
\usepackage{color}

\usepackage{fancyhdr}
\begin{document}

\title{Records in stochastic processes - Theory and applications}

\author{Gregor Wergen}

\address{Institut f\"ur Theoretische Physik, Universit\"at zu
K\"oln, 50937 K\"oln, Germany}
\ead{gw@thp.uni-koeln.de}
\begin{abstract} In recent years there has been a surge of interest in the statistics of record-breaking events in stochastic processes. Along with that, many new and interesting applications of the theory of records were discovered and explored. The record statistics of uncorrelated random variables sampled from time-dependent distributions was studied extensively. The findings were applied in various areas to model and explain record-breaking events in observational data. Particularly interesting and fruitful was the study of record-breaking temperatures and their connection with global warming, but also records in sports, biology and some areas in physics were considered in the last years. Similarly, researchers have recently started to understand the record statistics of correlated processes such as random walks, which can be helpful to model record events in financial time series. This review is an attempt to summarize and evaluate the progress that was made in the field of record statistics throughout the last years.
\end{abstract}

\maketitle

\setcounter{tocdepth}{2}
\tableofcontents
\pagestyle{fancy}
\fancyhead[L]{\textit{Records in stochastic processes - Theory and applications}}
\fancyhead[R]{\thepage}

\section{Why records?}

In our competitive society we care a lot about performance and we often feel the need to outperform others. Maybe this is why recently also researchers have become more and more interested in records. A record is simply an achievement, a result or some other kind of measurement in a given chain of events that exceeds everything that has been encountered previously. Therefore a new record is always something remarkable which attracts attention regardless of whether or not the occurrence of this record is considered good or bad. Records receive more attention and are remembered longer than other measurements because they show the boundary of what has been possible so far. In this context, the famous book 'Guinness World Records' holds its own record as the best-selling copyrighted book in history \cite{Guinness}.

An area where records are certainly of great interest is, of course, sports. Particularly in athletics and in swimming, Olympic- or world-records are always something special and noteworthy \cite{Gembris2002,Gembris2007}. But also in the context of global warming records have recently become particularly important and interesting for climatologist. The question how a changing climate affects the number of record temperatures that we encounter has bothered both the general public and researchers \cite{Stott2004,Benestad2003,Benestad2004,Redner2006,Meehl2009,Wergen2010,Newman2010,Elguindi2012,Rahmstorf2011}. By now it is well established that global warming leads to many new heat-records and to a decreased number of new record-breaking cold temperatures. 

Records are important also in countless other areas of science. In physics, they were discussed in the context of the theory of spin-glasses \cite{Jensen2005,Sibani2006,Sibani2007} and high-temperature superconductors \cite{Jensen2005,Oliveira2005}, but they also found applications in evolutionary biology \cite{Kaufmann1998,Sibani2004,Krug2005,Franke2011a}. Curiously, in 2010, the dynamics of ant movements were studied using results from the theory of records \cite{Richardson2010}. Thanks to new theoretical results it was recently possible to analyze and model the statistics of records in stock prices \cite{Wergen2011b,Majumdar2012,Wergeninprep}. 

These data-oriented studies were accompanied and complemented by a substantial number of new theoretical results. The classical theory of records in time series of independent and identically distributed (i.i.d.)~random variables was already developed many decades ago \cite{Arnold1998,Nevzorov2004,Glick1978}, but to understand the record statistics of more complicated systems such as the world's climate or evolutionary pathways, new techniques beyond this standard model were needed. In this context, various processes of uncorrelated random numbers sampled from time-dependent distributions were studied. Most importantly the Linear Drift Model (LDM), which was introduced already in the 80's, where random numbers are drawn from a distribution of unvarying shape but with an increasing mean value, was studied extensively \cite{Ballerini1985,Ballerini1987,Borovkov1999,Franke2010,Wergen2011,Franke2011b}. Some authors also considered record events from broadening distributions \cite{Krug2007,Eliazar2009}.

Also connected with problems in the adaptation of theoretical results from record statistics on observational data, are the so-called discreteness or rounding effects. Even though most of the classical theory is developed for random numbers sampled from continuous distributions, practical measurements are always imprecise and rounded to a certain accuracy. Both the record statistics of random numbers from discrete distributions \cite{Vervaat1973,Prodinger1996,Gouet2005,Gouet2007} as well as the consequences of analyzing records in time series of random numbers that were drawn from continuous distributions and then discretizes in a measuring process were discussed in recent years \cite{Wergen2012}.

In 2008, Majumdar and Ziff computed the record statistics of symmetric random walks \cite{Majumdar2008}. Their findings entailed a series of new theoretical results and applications. By now the complicated record statistics of biased random walks and L\'evy flights \cite{Wergen2011b,Majumdar2012} as well as the one of ensembles of multiple independent random walks \cite{Wergen2012a} is well understood. Records in continuous time random walks \cite{Sabhapandit2011} and also in the distance of higher dimensional jump processes from their origin \cite{Edery2011} were studied similarly.

The purpose of this work is to summarize and evaluate these recent developments mostly from a theoretical point of view, but also with a short evaluation of recent data-driven studies in the field. The rest of this review is organized as follows: We will start with a brief introduction to the classical theory of records, where we introduce the important notation and present some elementary results. Then, in section \ref{rev:uncorr}, we describe recent developments in the field of record statistics of uncorrelated random variables that are sampled from time-dependent distributions. In this context we consider the important Linear Drift Model of random variables with a linearly increasing mean value as well as a model of increasing variance. 

In the subsequent section \ref{rev:discreteness} we discuss various alternative models of records in continuous and discrete random variables. In particular, we will consider the effects of rounding and present generalized concepts like $\delta$- and geometric records, which are record events that are only counted if they exceed a certain barrier above or a certain multiple of the last record.

Then, in section \ref{rev:corr}, we analyze various stochastic processes with correlated entries starting with the symmetric random walk. After discussing the important results of Majumdar and Ziff on the symmetric discrete-time random walk \cite{Majumdar2008}, we will demonstrate how these findings can be generalized to biased random walks, to ensembles of multiple symmetric random walks and to symmetric continuous time random walks.

Various important applications that were mentioned above are presented and discussed in section \ref{rev:applications}. We will start by describing the progress made in the study of temperature records in \ref{rev:temperatures}. As an important application of the random walk model we outline some recent results about the statistics of record-breaking stock prices in section \ref{rev:finance}. Subsequently, we briefly mention some other applications, for instance in physics, biology and in athletics (\ref{rev:other_applications} and \ref{rev:athletics}). Afterwards, in section \ref{rev:summary}, we give a brief summary in which we assess the current state of research in the field of record statistics and point out a number of interesting open questions and suggestions for future research.

\section{Classical theory of records}
\label{rev:classical}

\begin{figure}[t]
\centerline{\includegraphics[width=0.75\textwidth]{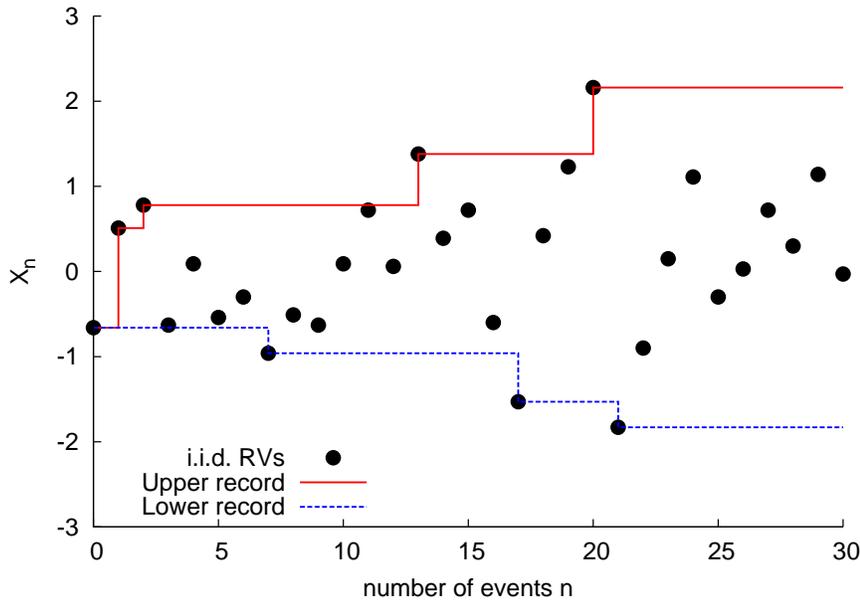}}
\caption{\label{rev:Fig_iid} Sketch of the record process of i.i.d.~RV's. The dots represent a time series $X_0,X_1,X_2,...$ of RV's drawn from a continuous distribution $f\left(x\right)$ (in this case a standard normal distribution). The red (blue dotted) lines illustrate the progressions of the upper (lower) record. Here, we find $5$ upper and $4$ lower records. In both cases $X_0$ is the first record. }
\end{figure}

Let us consider a time series $X_0,X_1,...,X_n$ of random variables (RV's), which can, for instance, be a series of temperatures, stock prices, sports results or some other kind of measurement process. In such a time series an entry $X_n$ is an upper record if it exceeds all previous entries:
\begin{eqnarray}
 X_n > \textrm{max}\{X_0,X_1,...,X_{n-1}\}.
\end{eqnarray}
Analogously, a lower record is an entry with $X_n<\textrm{min}\{X_0,X_1,...,X_{n-1}\}$. In general, one defines the first entry $X_0$ as the first (upper and lower) record. The record process in the simple case of independent and identically distributed (i.i.d.)~RV's is illustrated in Fig.~\ref{rev:Fig_iid}. Probably the two most studied quantities in the theory of records are the record number $R_n$ and the probability $P_n$ for a record at time $n$. This probability $P_n$ for an upper record is defined as 
\begin{eqnarray}
 P_n:=\textrm{Prob}\left[X_n > \textrm{max}\{X_0,X_1,...,X_{n-1}\}\right].
\end{eqnarray}
In the following we will also refer to $P_n$ as the record rate. The record number $R_n$ is simply the number of records that occurred in the time series up to time $n$. The mean record number $\langle R_n\rangle$, the expected average record number of a stochastic process, can by expressed in terms of the record rate:
\begin{eqnarray}
 \langle R_n\rangle = \sum_{k=0}^{n} P_k.
\end{eqnarray}
In the case of i.i.d.~RV's sampled from a continuous distribution with probability density function (pdf) $f\left(x\right)$, one can easily compute the record rate and the mean record number: With the so-called \textit{stick-shuffling} argument one finds that the probability $P_n$ for a record at time $n$ in a time series of i.i.d.~RV's is given by
\begin{eqnarray}\label{P_n_1_over_n}
 P_n = \frac{1}{n+1}.
\end{eqnarray}
This is just the probability that in a random ordering of $n+1$ RV's (sticks) the last one ($X_n$) is the largest. Despite the simplicity of this argument, it is important to notice that $P_n$ can also be computed more systematically. The probability for a RV sampled from $f\left(x\right)$ to be smaller than $x$ is clearly given by the cumulative distribution function (cdf) $F\left(x\right):=\int^x \mathrm{d}x\;f\left(x\right)$. Therefore, in the i.i.d. case, $P_n$ is given by the following integral:
\begin{eqnarray}
 P_n = \int \mathrm{d}x\; f\left(x\right)  F^{n}\left(x\right).
\end{eqnarray}
Here, the probability for a record with value $x$ at time $n$ is integrated over all possible values of $x$ to obtain $P_n$. Now, partial integration leads to
\begin{eqnarray}
 P_n = 1-n \int \mathrm{d}x\; f\left(x\right) F^{n}\left(x\right) = 1 - nP_n. 
\end{eqnarray}
and therefore $P_n=1/\left(n+1\right)$ as in Eq. \ref{P_n_1_over_n}.

With this result for the record rate, the mean record number $\langle R_n\rangle$ in a series of i.i.d.~RV's takes the form
\begin{eqnarray}
 \langle R_n\rangle = \sum_{k=0}^{n} P_k = \sum_{k=0}^n \frac{1}{k+1} = H_{n+1} \xrightarrow{n\rightarrow\infty} \ln n + \gamma,
\end{eqnarray}
where $H_k$ is the $k$th Harmonic number (cf. \cite{Abramowitz1970}) and $\gamma\approx0.577215...$ the Euler-Mascheroni constant \cite{Abramowitz1970, Arnold1998}.

An important feature of record events in i.i.d.~RV's is that they are stochastically independent. The probability for a record in the $n$th event is independent from records in previous entries \cite{Arnold1998,Franke2010}. One can shown that the joint probability $P_{n,m}$ of records both at times $n$ and $m$ factorizes \cite{Arnold1998}: 
\begin{eqnarray}
P_{n,m} := \textrm{Prob}\left[X_n,X_m\;\textrm{both records}\right] = P_n\cdot P_m 
\end{eqnarray}
For arbitrary $n$ and $m$, it is not straightforward to prove this result and we refer the reader to the book by Arnold et al.~\cite{Arnold1998}. However, for neighboring entries $n$ and $m=n+1$ it is possible to compute $P_{n,n+1}$ from the integral
\begin{eqnarray}
 P_{n,n+1} = \int \mathrm{d}x_{n+1}\;f\left(x_{n+1}\right) \int^{x_{n+1}} \mathrm{d}x_n\; f\left(x_n\right) F^n\left(x_n\right).
\end{eqnarray}
This is just the probability for a record at time $n+1$ with value $x_{n+1}$ and a previous record at time $n$ with value $x_n<x_{n+1}$ integrated over all possible values of $x_{n+1}$ and $x_n<x_{n+1}$. Again, this integral can be evaluated by elementary means and we find
\begin{eqnarray}
 P_{n,n+1} = \frac{1}{n+1}\frac{1}{n+2} = P_n P_{n+1}.
\end{eqnarray}

By now, a lot more is known about the record statistics of i.i.d.~RV's (while the main purpose of this review is to discuss the record statistics of time series of time-dependent and correlated RV's). A good review can be found in the book by Arnold et al.~\cite{Arnold1998}, or Nevzorov \cite{Nevzorov2004} (see also \cite{Glick1978}). There, quantities like the distributions of record values with a given record number, or the interesting waiting-time statistics between individual record events are discussed in detail. A noteworthy finding is that the mean time $\langle T_{R_n}\rangle$, at which a record with record number $R_n$ occurs is infinite (see also \cite{Shorrock1972a,Shorrock1972b}). Similarly the inter-record times $\Delta_{R_n}:= T_{R_n}-T_{R_{n-1}}$ have a divergent mean value $\langle \Delta_{R_n}\rangle$.

Furthermore, in the book by Arnold et al.~\cite{Arnold1998}, it is shown how to compute the probability density function of a record value with a given record number $k$. Arnold et al.~\cite{Arnold1998} consider the the pdf $f_k\left(x\right)$ of all record values with a fixed record number $k$, where these records can occur at an arbitrary time $n$. He argues that due to the so-called \textit{lack-of-memory} property of the exponential distribution with $f\left(x\right) = e^{-x}$ (for $x>0$), the pdf $f_k\left(x\right)$ of a record value with the record number $k$ from the exponential distribution is given by
\begin{eqnarray}\label{rev:f_k_x}
 f_k\left(x\right) = \frac{1}{\left(k-1\right)!} x^{-k} e^{-x},
\end{eqnarray}
Since the RV's larger than a value $\tilde{x}>0$ sampled from $f\left(x\right) = e^{-x}$ are again exponentially distributed (with pdf $e^{-\left(x-\tilde{x}\right)}$ for $x>\tilde{x}$), a record with record number $k$ from an exponential distribution is given by the value of the $\left(k-1\right)$th record plus an exponential RV sampled from $f\left(x\right)$. Therefore, the pdf of the $k$th record is just the convolution of $f_{k-1}\left(x\right)$ and $f\left(x\right)$. By iteration this leads eventually to the Gamma-distribution in Eq.~\ref{rev:f_k_x}.

This result can be used to compute the distribution of the $k$th record value in time series of RV's from arbitrary continuous distributions. For that purpose it is useful to express the distribution of a RV $X_i$ from an arbitrary continuous pdf $f\left(x\right)$ in terms of the exponential distribution. $X_i$ can be expressed by an exponential RV $X_i^{\textrm{exp}}$ as follows:
\begin{eqnarray}\label{rev:F_inverse_exp}
 F^{-1}\left(1-\textrm{exp}\left(-X_i^{\left(\textrm{exp}\right)}\right)\right),
\end{eqnarray}
where $F^{-1}\left(x\right)$ is the inverse cumulative of $f\left(x\right)$ and $X_i^{\left(\textrm{exp}\right)}$ an exponentially distributed RV with pdf $e^{-x}$ as before \cite{Arnold1998}. Since $1-\textrm{exp}\left(-X_i^{\left(\textrm{exp}\right)}\right)$ is just a uniform distribution on the interval $\left[0,1\right)$, which is the image space of the cdf $F\left(x\right)$, Eq. \ref{rev:F_inverse_exp} must be distributed according to $F\left(x\right)$. Using Eq. \ref{rev:F_inverse_exp} we can infer that
\begin{eqnarray}
 F\left(x\right) = \textrm{Prob}\left[X_i < x\right] & = & \textrm{Prob}\left[F^{-1}\left(1-e^{-X_i^{\left(\textrm{exp}\right)}}\right) < x\right] \nonumber \\
 & = & \textrm{Prob}\left[X_i^{\left(\textrm{exp}\right)} < -\ln \left(1-F\left(x\right)\right)\right].
\end{eqnarray}
With this result it is clear how to compute $f_k\left(x\right)$ in the general case. We just have to replace the $x$ in Eq. \ref{rev:f_k_x} by $-\ln \left(1-F\left(x\right)\right)$. This leads to
\begin{eqnarray}
 f_k\left(x\right) = \frac{1}{\left(k-1\right)!} \left(-\ln \left(1-F\left(x\right)\right)\right)^k f\left(x\right).
\end{eqnarray}



\bigskip
An important results in extreme value statistics is that the distribution of the maximum
\begin{eqnarray}
 M_n := \textrm{max}\{X_0,X_1,...,X_n\}
\end{eqnarray}
of a given set $X_0,X_1,...,X_n$ of i.i.d.~RV's sampled from a continuous pdf $f\left(x\right)$ converges to one of three possible limiting distributions \cite{Gumbel1954} (for a detailed introduction see the books by Galambos \cite{Galambos1987} or De Haan and Ferreira \cite{DeHaan2006}). According to the celebrated Fisher-Tippett-Gnedenko Theorem \cite{Galambos1987,DeHaan2006} of extreme value theory, the limiting distribution of $M_n$ can always be rescaled to one of the following shapes:

\paragraph{I - Weibull distribution:} For RV's with a finite support, the rescaled maximum approaches a (reversed) Weibull distribution with the cdf
\begin{eqnarray}
 F_{\mathrm{I}}\left(x\right) = \begin{cases} e^{-\left(-x\right)^{-\kappa}},& \textrm{for }x<0,\\
 1, &\textrm{for }x\ge 0 \end{cases},
\end{eqnarray}
where, $\kappa<0$ is a free paramter.

\paragraph{II - Gumbel distribution:} The (rescaled) maximum of RV's from distributions with an infinite support decaying faster than a power-law such as, for instance, with an exponential tail, is distributed according to a Gumbel distribution:
\begin{eqnarray}
 F_{\mathrm{II}}\left(x\right) = e^{e^{-x}}.
\end{eqnarray}

\paragraph{III - Fr\'echet distribution:} For RV's with an infinite support and power-law tails, the (rescaled) distribution of the maximum converges to the Fr\'echet distribution:
\begin{eqnarray}
 F_{\mathrm{III}}\left(x\right) = \begin{cases} 0, & \textrm{for }x<0,\\ e^{-x^{-\kappa}}, & \textrm{for }x\geq0 \end{cases}
\end{eqnarray}
with a free parameter $\kappa>0$.

It turns out that these universality classes are also relevant for the distributions of record values. In fact, in 1973, Resnick~\cite{Resnick1973} could prove that the distributions of record values with a record number $k$ converge to a limit law of the form:
\begin{eqnarray}
 \Phi \left(-\ln\left(-\ln\left(F_i\left(x\right)\right)\right)\right),
\end{eqnarray}
where the functions $F_i\left(x\right)$ with $i=\mathrm{I},\mathrm{II},\mathrm{III}$ are the limiting distributions of the maximal value ($\mathrm{I}$ - Weibull, $\mathrm{II}$ - Gumbel, $\mathrm{III}$ - Fr\'echet) and $\Phi\left(x\right)$ is a Gaussian cdf with
\begin{eqnarray}
 \Phi\left(x\right) = \frac{1}{2\pi} \int_{-\infty}^x e^{-t^2/2} \mathrm{d}t.
\end{eqnarray}
Using this result Resnick could show that the (rescaled) limiting distribution of a record with given record number $k$ can only approach one of these three limiting forms \cite{Resnick1973}:

\paragraph{I - Negative-log-normal distribution:} For RV's from a distribution with a finite support (Weibull class), the distribution of the (rescaled) record values approaches the following Negative-log-normal form:
\begin{eqnarray}
 \Phi_{\mathrm{I}}\left(x\right) = \begin{cases} \Phi\left(\ln\left(-x\right)^{-\kappa}\right), & \textrm{for }x<0,\\ 1, & \textrm{for }x\geq0 \end{cases}
\end{eqnarray}
with $\kappa<0$.

\paragraph{II - Normal distribution:} Record values from time series of RV's from the Gumbel class are asymptotically normal distribution:
\begin{eqnarray}
 \Phi_{\mathrm{II}}\left(x\right) = \Phi\left(x\right).
\end{eqnarray}

\paragraph{III - Log-normal distribution:} The record values of RV's of the Fr\'echet type have a rescaled limiting distribution of the Log-normal form
\begin{eqnarray}
 \Phi_{\mathrm{III}}\left(x\right) = \begin{cases} 0,& \textrm{for }x<0,\\
 \Phi\left(\ln \left(x^{-\kappa}\right)\right), &\textrm{for }x\ge 0 \end{cases}
\end{eqnarray}
with $\kappa>0$.

In the following, we will find that the three universality classes of EVS are also of importance for the record statistics of time-dependent RV's. Many of the results presented in this article will be characterized and discussed in the context of these classes. In time series of correlated RV's however, as we will see in section \ref{rev:corr}, the classes lose their importance and one finds different interesting universal characteristics.

\section{Records in uncorrelated and time-dependent RV's}
\label{rev:uncorr}

While the classical theory introduced above deals with identically distributed RV's drawn from a single, stationary pdf $f\left(x\right)$, one can also consider the more general scenario of uncorrelated, but non-identically distributed random numbers $X_0,X_1,...,X_n$ from a time series of probability densities $f_i\left(x_i\right)$. In this general case, it is more complicated to compute the record rate $P_n$ and the mean record number $\langle R_n\rangle$. Here, the record rate can be obtained from the integral (cf.~\cite{Franke2010})
\begin{eqnarray}\label{rev:P_n_general}
 P_n = \int\mathrm{d}x_n\;f_n\left(x_n\right)\prod_{k=0}^{n-1} F_k\left(x_n\right),
\end{eqnarray}
where $F_k\left(x_n\right) = \int^x_n \mathrm{d}x_k\; f_k\left(x_k\right)$ is the cumulative distribution function (cdf) of the pdf $f_k\left(x_k\right)$. Eq. \ref{rev:P_n_general} can be understood as follows: $f_n\left(x_n\right)$ gives the probability that the $n$th RV assumes the value $x_n$. This is multiplied with the probability $\prod_{k=0}^{n-1} F_k\left(x_n\right)$ that all previous RV's $X_0,...,X_{n-1}$ are smaller than $x_n$. Therefore $f_n\left(x_n\right)\prod_{k=0}^{n-1} F_k\left(x_n\right)$ is the probability that the $n$th RV is a record with value $x_n$ and $P_n$ can be obtained by integrating over all possible values of $x_n$.

In the following we present two possible choices for the series of probability densities $f_k\left(x_k\right)$ that were studied in the literature and that proved to be useful in the analysis of observational data.

\subsection{The Linear Drift Model}
\label{rev:sec_ldm}

\begin{figure}[t]
\centerline{\includegraphics[width=0.75\textwidth]{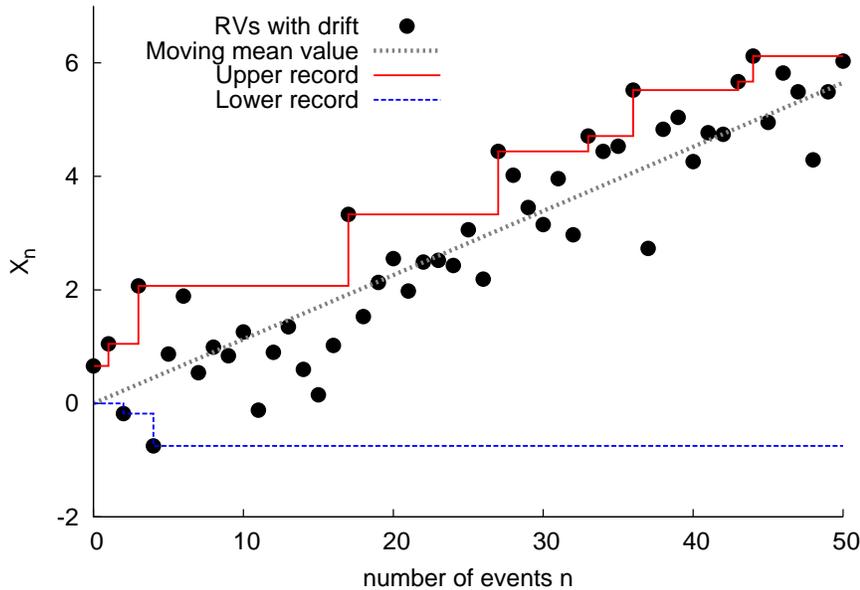}}
\caption{\label{rev:Fig_LDM} Sketch of the record process of uncorrelated RV's with a linear drift ($c=0.1$). The dots represent a time series $X_0,X_1,X_2,...$ of RV's drawn from a series continuous distributions with $f_k\left(x\right) = f\left(x-ck\right)$. The red (blue dotted) lines illustrate the progressions of the upper (lower) records. }
\end{figure}

The Linear Drift Model (LDM) was first introduced by Ballerini and Resnick in the 1980's \cite{Ballerini1985,Ballerini1987} and later studied by Borovkov \cite{Borovkov1999} and more recently by Franke et al.~\cite{Franke2010} as well as Wergen et al.~\cite{Wergen2011}. The model describes RV's drawn from a distribution that retains its shape but has a time-dependent mean value. In particular, the RV's are sampled from a series of pdf's $f_k\left(x_k\right) = f\left(x_k-ck\right)$, where $c$ is a real constant, which is called the drift. The entries of such a time series are of the form
\begin{eqnarray}
 X_k = Y_k + ck,
\end{eqnarray}
where $Y_0,Y_1,...,Y_n$ is a time series of i.i.d.~RV's sample from $f\left(x\right)$. The record process of a series of RV's from the LDM is illustrated in Fig. \ref{rev:Fig_LDM}. Here, the general, time-dependent expression for the record rate (Eq.~\ref{rev:P_n_general}) takes the following form:
\begin{eqnarray}
 P_n\left(c\right) & = & \int\mathrm{d}x\;f\left(x-cn\right) \prod_{k=0}^{n-1} F\left(x-ck\right) \nonumber\\
 & = & \int\mathrm{d}x\;f\left(x\right) \prod_{k=1}^{n} F\left(x+ck\right)
\end{eqnarray}
Ballerini and Resnick \cite{Ballerini1985} could prove that this record rate has an asymptotically constant limiting value $P\left(c\right) = \lim_{n\rightarrow\infty}P_n\left(c\right)$ if the pdf $f\left(x\right)$ has a finite first moment $\int \mathrm{d}x\; xf\left(x\right) < \infty$. To determine the behavior of $P_n\left(c\right)$ in more detail, is however a difficult problem, which, in general, can not be solved exactly.

There is an example of a pdf, for which the record rate $P_n\left(c\right)$ in the LDM can be calculated for arbitrary $n$: For the Gumbel distribution with the probability density $f\left(x\right) = e^{-x}\textrm{exp}\left(e^{-x}\right)$, Franke et al.~\cite{Franke2010} found that
\begin{eqnarray}
 P^{\textrm{Gumbel}}_n\left(c\right) = \frac{1-e^{-c}}{1-e^{-nc}}
\end{eqnarray}

Similarly, it is possible to compute the asymptotic record rate for the exponential distribution with $f\left(x\right) = \nu e^{-\nu x}$ (with $x>0$ and $\nu>0$). In this case, the record rate $P_n\left(c\right)$ assumes the following form:
\begin{eqnarray}\label{rev:Pnc_exp}
 P_n\left(c\right) & = & \int_0^{\infty} \mathrm{d}x\; \nu e^{-\nu x} \prod_{k=1}^n \left(1-e^{-\nu \left(x+ck\right)}\right) \nonumber \\
  & = & \int_0^1 \mathrm{d}y\; \frac{ \left(y,e^{-c\nu} \right)_n }{1-y},
\end{eqnarray}
where $\left(a,q\right)_n$ is the q-Pochhammer symbol with $\left(a,q\right)_n:=\prod_{k=0}^n \left(1-aq^k\right)$ \cite{Abramowitz1970}. With this we can expand the asymptotic record rate $P\left(c\right)$ in powers of $e^{-c\nu}$ and find
\begin{eqnarray}\label{rev:Pc_exp}
 P\left(c\right) & = & \int_0^1 \mathrm{d}y\; \frac{ \left(y,e^{-c\nu} \right)_{\infty} }{1-y} \nonumber \\
  & \approx & 1 - \frac{1}{2}e^{-c\nu} - \frac{1}{2}e^{-2c\nu} - \frac{1}{6}e^{-3c\nu} - \frac{1}{6}e^{-4c\nu} + O\left(e^{-5c\nu}\right).
\end{eqnarray}
By using computer algebra software, such as Mathematica, it is possible to compute arbitrary higher order terms of this expansion. However, we found that, in comparison with numerical simulations of $P\left(c\right)$, the expansion up to the 4th order in Eq.~\ref{rev:Pc_exp} is already very accurate and fails only for small $c\rightarrow 0$ \cite{WergenUnpup}. For this case, one can compute the record rate $P_n\left(c\right)$ by a different approach. Replacing the product in Eq.~\ref{rev:Pnc_exp} by the exponential of a sum of logarithms leads to
\begin{eqnarray}
 P\left(c\right) & = & \int_0^{\infty} \mathrm{d}x\; \nu e^{-\nu x} \textrm{exp}\left(-\nu x-\sum_{k=1}^n \ln \left(1-e^{-\nu\left( x-ck\right)}\right)\right) \nonumber \\ & \approx & \int_0^{\infty} \mathrm{d}x\; \textrm{exp} \left(\frac{e^{-\nu x}}{c\nu}\left(1-e^{-c\nu n}\right)\right),
\end{eqnarray}
where, for the second step, we replaced the sum by an integral assuming that $n\gg1$ and $c\nu\ll1$. With this we obtain the small $c$ behavior of $P_n\left(c\right)$ for the exponential case:
\begin{eqnarray}\label{rev:Pnc_exp_smallc}
 P_n\left(c\right) \approx c\nu\frac{1-e^{-\frac{1}{c\nu}\left(1-e^{-c\nu n}\right)}}{1-e^{-c\nu n}} \xrightarrow[n\rightarrow\infty]{} c\nu\left(1-e^{-\frac{1}{c\nu}}\right),
\end{eqnarray}
which, for small $c\nu\ll1$, approaches $c\nu$. Apparently, for small $c$, the record rate of the exponential distribution depends linearly on $c$. Comparing with numerical simulations, we found that Eq. \ref{rev:Pnc_exp_smallc} computes $P_n\left(c\right)$ accurately for $c\nu<\frac{1}{2}$ \cite{WergenUnpup}.

In the article by Franke et al.~\cite{Franke2010}, the record rate for a more general set of continuous probability distributions was computed in two different regimes. For small $c$, Franke et al. derived approximate results for finite values of $n$ in the regime of $nc\ll\sigma$, where $\sigma$ is usually the standard deviation or some other measure of the width of $f\left(x\right)$. The same can be done in the opposite regime of $c\rightarrow\infty$. It turns out that the behavior of $P_n\left(c\right)$ depends systematically on the three classes of EVS. Franke et al.~\cite{Franke2010} discussed their findings in the context of these classes.

\subsubsection{The regime of small $cn$}

In the small $c$ regime, it is possible to expand the record rate $P_n\left(c\right)$ into powers of $c$. Expansion up to the first order yields
\begin{eqnarray}\label{rev:derivation_p_n_c}
 P_n\left(c\right) & = & \int \mathrm{d}x\; f\left(x\right) \prod_{k=1}^{n} F\left(x+ck\right) \nonumber \\
   & \approx & \int \mathrm{d}x\; f\left(x\right) \prod_{k=1}^{n}\left(F\left(x\right) + ckf\left(x\right)\right) \nonumber \\
   & \approx & \int \mathrm{d}x\; f\left(x\right) F^{n}\left(x\right) + \frac{c}{2}n\left(n+1\right)\int \mathrm{d}x\; f^2\left(x\right) F^{n-1}\left(x\right).
\end{eqnarray}
The first summand in the last line is the stationary record rate with $P_n\left(c=0\right)=1/n$. With
\begin{eqnarray}
 \textrm{I}_n := \int \mathrm{d}x\; f^2\left(x\right) F^{n-1}\left(x\right)
\end{eqnarray}
this leads to
\begin{eqnarray}
 P_n\left(c\right) \approx \frac{1}{n+1} + \frac{c}{2}n\left(n+1\right) \textrm{I}_n.
\end{eqnarray}
This expansion is accurate if the underlying distribution $f\left(x\right)$ varies only slowly between $x$ and $x+cn$. For many probability densities this can be translated into $cn\ll\sigma$ (with $\sigma^2 := \int \mathrm{d}x\; x^2 f\left(x\right)$).

In \cite{Franke2010}, $\textrm{I}_n$ was computed for several representative distributions from the three classes of extreme value statistics. Here, in order to get a clear picture of the effect of the drift on the record rate depending on the extreme value class of the underlying distribution, we want to derive $\textrm{I}_n$ and $P_n\left(c\right)$ for a Generalized Pareto Distribution (GPD). We consider RV's from the cdf
\begin{eqnarray}\label{rev:generalized_pareto}
 F_{\xi}\left(x\right) = \begin{cases} 1-\left(1+\xi x\right)^{-\frac{1}{\xi}}, & \textrm{for }\xi\neq 0\\
            1-e^{-x}, & \textrm{for }\xi= 0.
           \end{cases}
\end{eqnarray}
$\xi\in\mathbb{R}$ is the shape parameter of $F\left(x\right)$. For $\xi\geq0$ this distribution has an infinite support and is defined for $x>0$, for $\xi<0$ is it defined on the finite interval $x\in\left[1,1-\xi^{-1}\right]$.

Depending on $\xi$, the GPD can be in all three classes of EVS. For $\xi<0$, $F_{\xi}\left(x\right)$ is in the Weibull class, for $\xi=0$ in the Gumbel class and for $\xi>0$ in the Fr\'echet class.

For this distribution, the integral $\textrm{I}_n$ can be evaluated by elementary means and we find that
\begin{eqnarray}
 \textrm{I}_n = \begin{cases} \xi \textrm{B}\left(n,2+\xi\right), & \textrm{for }\xi\neq 0\\
               1 / \left(n\left(n+1\right)\right), & \textrm{for }\xi= 0,
              \end{cases}
\end{eqnarray}
where  $\textrm{B}\left(x,y\right)=\Gamma\left[x\right]\Gamma\left[y\right]/\Gamma\left[x+y\right]$ is the Beta-function \cite{Abramowitz1970}. Using Stirling's approximation \cite{Abramowitz1970}, we find that for $n\gg1$, the record rate of the GPD with a drift $c\ll n^{\xi-1}$ is given by
\begin{eqnarray}
 P_n\left(c\right) \approx \frac{1}{n+1} + c\;\begin{cases} \xi \Gamma\left[2+\xi\right] n^{-\xi}, & \textrm{for }\xi\neq 0 \\ 
 \frac{1}{2}, & \textrm{for }\xi= 0. \end{cases}
\end{eqnarray}
This result summarizes how a small linear drift affects the record rates depending on the extreme value class of the underlying distribution. Although this is no conclusive proof, we conjecture that the effect of the drift generally increases with $n$ for distributions of the Weibull class. In the Fr\'echet class the effect decays with $n$ and the drift is asymptotically negligible. The Gumbel class is intermediate between these two cases. Interestingly, since for $\xi>1$, $n^{\xi-1}$ grows with $n$, some of the results for the Fr\'echet class (for $\xi>1$) are also correct in the asymptotic limit with $n\rightarrow\infty$.

To better understand the behavior of $P_n\left(c\right)$ in the Gumbel class, Franke et al.~\cite{Franke2010} considered the Generalized Gaussian Distribution (GGD) with
\begin{eqnarray}
f\left(x\right) = 2\Gamma\left[1+\beta^{-1}\right]^{-1}e^{-|x|^{\beta}}
\end{eqnarray}
with $\beta>0$. They could show that, here, $\textrm{I}_n$ grows logarithmically with $n$:
\begin{eqnarray}
 \textrm{I}_n \propto \ln \left(n\right)^{1-\frac{1}{\beta}}.
\end{eqnarray}
This expression includes the important case of a Gaussian distribution for $\beta=2$. For the Gaussian probability density $f\left(x\right) = \frac{1}{\sqrt{2\pi}}e^{-x^2/2\sigma^2}$ one obtains
\begin{eqnarray}\label{rev:ldm_Gaussian}
 P_n\left(c\right) \approx \frac{1}{n+1} + \frac{c}{\sigma}\frac{2\sqrt{\pi}}{e^2}\sqrt{\ln \left(\frac{n^2}{8\pi}\right)}.
\end{eqnarray}

\subsubsection{Correlations in the Linear Drift Model}

An interesting subtlety that was discovered in the study of the LDM is the fact that record events in this process are not stochastically independent as in the i.i.d.~case. In particular, it was found by Wergen et al.~\cite{Wergen2011} that the probability $P_{n,m}\left(c\right)$ of records in the entries $n$ and $m$ in a series of RV's with a linear drift can differ from the product of the record rates $P_n\left(c\right)$ and $P_m\left(c\right)$. In \cite{Wergen2011}, the probability $P_{n,n+1}\left(c\right)$ of having two consecutive records was studied in detail. They showed that, depending on the choice of the underlying distribution, $P_{n,n+1}\left(c\right)$ can be both, smaller and larger than $P_n\left(c\right)\cdot P_{n+1}\left(c\right)$. Therefore, the probability for a second record in step $n+1$ after a record in step $n$ can be both increased and decreased with respect to the unconditional probability $P_{n+1}\left(c\right)$.

Wergen et al.~\cite{Wergen2011} defined the ratio
\begin{eqnarray}
 l_{n,n+1}\left(c\right) := \frac{P_{n,n+1}\left(c\right)}{P_n\left(c\right)\cdot P_{n+1}\left(c\right)},
\end{eqnarray}
which is always given by $l_{n,n+1}\left(c=0\right)=1$ in the i.i.d.~case. In the regime of small $c$ and $n\gg1$, we can again use the GPD (Eq. \ref{rev:generalized_pareto}) to illustrate the asymptotic behavior of this quantity. Expanding $l_{n,n+1}\left(c\right)$ up to first order in $c$ with the same method as before (see the derivation following Eq. \ref{rev:derivation_p_n_c}), we find that
\begin{eqnarray}
 l_{n,n-1} \approx 1 + c \begin{cases} \xi \Gamma\left[2+\xi\right] n^{1-\xi}, & \textrm{for }\xi\neq 0 \\ 
 \frac{1}{2}, & \textrm{for }\xi= 0, \end{cases}
\end{eqnarray}
which is again valid for $c\ll n^{\xi-1}$. Apparently, the exponential distribution with $f\left(x\right) = e^{-x}$ ($\xi = 0$) plays an outstanding role. For the distributions of the Weibull class with $\xi<0$, the inter-record correlations are negative for a positive drift $c>0$, for the representatives of the Fr\'echet class ($\xi>0$), $l_{n,n-1}\left(c>0\right)$ is larger than one and grows with $n$. Only in the exponential case, $l_{n,n+1}\left(c\right)$ assumes an $n$-independent value slightly above unity. 

Again, the intermediate Gumbel regime can be studied more systematically using the GGD with $f\left(x\right) \propto e^{-|x|^{\beta}}$ as before. In a lengthy calculation, Wergen et al.~\cite{Wergen2011} showed that, here, for $n\gg1$, the ratio $l_{n,n+1}\left(c\right)$ behaves like
\begin{eqnarray}
 l_{n,n+1}\left(c\right) \approx 1 - cnA\left(1-\frac{1}{\beta}\right) \ln \left(n\right)^{1-\frac{1}{\beta}}.
\end{eqnarray}
with a positive constant $A$, which depends on $n$. Apparently, for $\beta<1$, stretched exponential distributions, broader than the exponential have positive correlations that grow logarithmically with $n$. Distributions decaying faster than the exponential ($\beta>1$) lead to negative correlations.

Even though it is not clear how to explain the emergence of these correlations and, in particular, why it is possible that records occur more frequently after a preceding record, these effects turn out to be useful. Since it is a well known problem for an experimentalist to decide whether or not a series of measurements is drawn from an underlying distribution with so-called heavy-tails (see for instance \cite{albatros,albatros2,Clauset2009}), Franke et al.~\cite{Franke2011b} proposed a test that uses the findings presented in \cite{Wergen2011} in this matter.

For this test a set of measurements $X_0,X_1,...,X_n$ has to be shuffled randomly before adding an artificial linear drift. For a random permutation $\pi_0,\pi_1...,\pi_n$ of $0,1,...,n$ such a shuffled and artificially drifted set of data is given by
\begin{eqnarray}
 X_{\pi_0},X_{\pi_1}+c,X_{\pi_2}+2c,...,X_{\pi_n}+nc.
\end{eqnarray}
Now one can analyze the inter-record correlations in this time series. In particular, one has to compute the ratio $l_{n,n+1}\left(c\right)$. As shown in \cite{Franke2011b}, the statistics can be improved significantly by averaging over many different random permutations. If the correlations in the drifted time series are positive ($l_{n,n+1}\left(c\right)>1$) this is a good indicator for measurements from a distribution, which is at least broader than the exponential one. Franke et al.~\cite{Franke2011b} demonstrated that this record-based test allows one to detect these heavy-tail properties already in very small data-sets with less than $64$ data-points. In this context, the test might be better than standard methods like, for instance, maximum likelihood estimators, which are commonly used for problems of this type. However, a thorough comparison of this new test with the existing ones has not been performed yet.

\subsection{The Increasing Variance Model}

\begin{figure}[t]
\centerline{\includegraphics[width=0.75\textwidth]{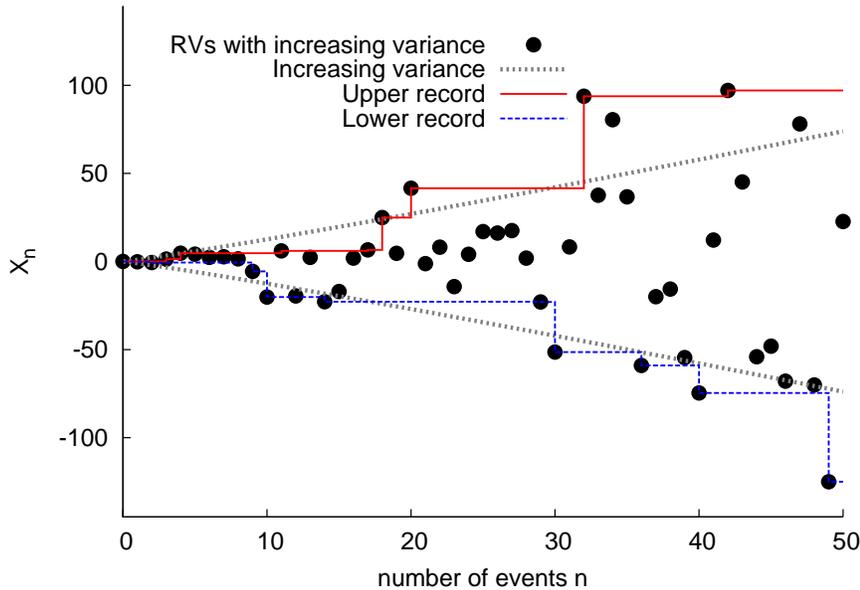}}
\caption{\label{rev:Fig_IVM} Sketch of the record process of uncorrelated (Gaussian) RV's from a broadening distribution. The dots represent a time series $X_0,X_1,X_2,...$ of RV's drawn from a series of continuous and symmetric distributions $f_k\left(x\right) = f\left(xk^{-\alpha}\right)$ (Here: $\alpha=1.1$). The red (blue dotted) lines illustrate the progression of the upper (lower) record.}
\end{figure}

In 1975, Yang \cite{Yang1975} introduced a model of growing populations to explain the increased record rate in sports due to a growing number of athletes who attempt to break records. Yang showed that any exponentially growing population of athletes leads to an asymptotically constant record rate $\lim_{n\rightarrow\infty} P_n>0$.

Building up on this model, Krug \cite{Krug2007} considered random variables $X_0,X_1,...,X_n$ from a series of probability densities $f_k\left(x_k\right)$ with a time-dependent width:
\begin{eqnarray}
 f_k\left(x_k\right) = \lambda_k f\left(\lambda_k x\right).
\end{eqnarray}
In particular, Krug discussed distributions with a power-law time-dependence and 
\begin{eqnarray}
 \lambda_k = k^{-\alpha}
\end{eqnarray}
Such a process is illustrated in Fig. \ref{rev:Fig_IVM} for an $\alpha$ slightly larger than one. Clearly, the distribution broadens for $\alpha>0$ and gets sharper when $\alpha<0$. Here, the record rate $P_n\left(\alpha\right)$ takes the following form:
\begin{eqnarray}\label{rev:ivm_general}
 P_n\left(\alpha\right) & = & \int \mathrm{d}x\; f\left(xn^{-\alpha}\right) \prod_{k=0}^{n-1} k^{-\alpha} F\left(x k^{-\alpha}\right) \nonumber \\ & = & \int \mathrm{d}x\; f\left(x\right) \prod_{k=0}^{n-1} F\left(x \left(\frac{k}{n}\right)^{-\alpha}\right) 
\end{eqnarray}
Krug \cite{Krug2007} computed the asymptotic behavior of the record rate $P_n\left(\alpha\right)$ and the mean record number $\langle R_n\left(\alpha\right)\rangle$ for this model in the context of the three universality classes of EVS. The effect of a broadening distribution with $\alpha>1$ is similar to the effect of a positive drift in the LDM. For distributions of the Fr\'echet class, the broadening width does not systematically change the large $n$ behavior of the record rate. At the same time, it has the strongest effects in the Weibull class.

Using the findings of Krug \cite{Krug2007}, we can calculate the asymptotic behavior of the record rate $P_n\left(\alpha\right)$ for the GPD $F_{\xi}\left(x\right)$, which was introduced in section \ref{rev:sec_ldm}. For a large $n\rightarrow\infty$ and $\alpha>1$ we obtain
 \begin{eqnarray}
 P_n\left(\alpha\right) \propto \begin{cases} \alpha^{\left(1-\xi\right)^{-1}} n^{-\left(1-\xi\right)^{-1}}, & \textrm{for } \xi < 0 \quad \textrm{(Weibull class)} \\ \frac{\ln\left(n\right)}{n} & \textrm{for } \xi = 0 \quad \textrm{(Exp. distribution)} \\ \frac{1}{n} & \textrm{for } \xi > 0 \quad \textrm{(Fr\'echet class)} \end{cases}
\end{eqnarray}
For the mean record number, this leads to
\begin{eqnarray}
 \langle R_n\left(\alpha\right)\rangle \propto \begin{cases} \alpha^{\left(1-\xi\right)^{-1}} n^{-\xi\left(1-\xi\right)^{-1}}, & \textrm{for } \xi < 0 \quad \textrm{(Weibull class)} \\ \left(\ln\left(n\right)\right)^2 & \textrm{for } \xi = 0 \quad \textrm{(Exp. distribution)} \\ \ln\left(n\right) & \textrm{for } \xi > 0 \quad \textrm{(Fr\'echet class)} \end{cases}
\end{eqnarray}

For a Gaussian distribution, the asymptotic results only differ in the prefactors from the exponential case.

Krug also studied the correlations between the record events in this model in a numerical manner. In contrast to the LDM, he found only negative correlations between records from RV's with an increasing variance. As in the case of the LDM, it is still controversial how to explain these correlations comprehensibly.

\section{Discreteness, rounding and ties}
\label{rev:discreteness}

The theoretical results in the previous chapter were all derived for RV's from entirely continuous distributions. In the context of experimental measurements and their record statistics, but also for purely mathematical reasons, one can also study models with discrete RV's with respect to records. In this case the statistics of records is more complicated and, in principle, there are several different approaches to this problem.

On the one hand, it is possible to consider the record statistics of RV's from distributions which are inherently discrete. Two prominent examples are 
\begin{itemize}
\item the discrete uniform distribution with equally likely probabilities for a finite number of RV's: $P\left[X=k\right] = 1/N$ (with $N\in\mathbb{N}$ being the number of possible outcomes and $k=1,...,N$), 
\item and the geometric distribution with $P\left[X=k\right] = (1-p)^{k-1}p$ (with $p\in\left[0,1\right]$ and $k\in\mathbb{N}$.)
\end{itemize}
In the case of a discrete distribution, it is possible that a record value, for instance an entry $X_i$, gets tied by a succeeding RV $X_j=X_i$. This is impossible for RV's sampled from a continuous pdf. In the case of a tie, one has to decide whether or not one wants to count this tie as a new record. In the literature about record statistics from discrete distributions, records without ties are usually called \textit{strong} records, while records including ties are called \textit{weak} records.

\begin{figure}[t]
\centerline{\includegraphics[width=0.75\textwidth]{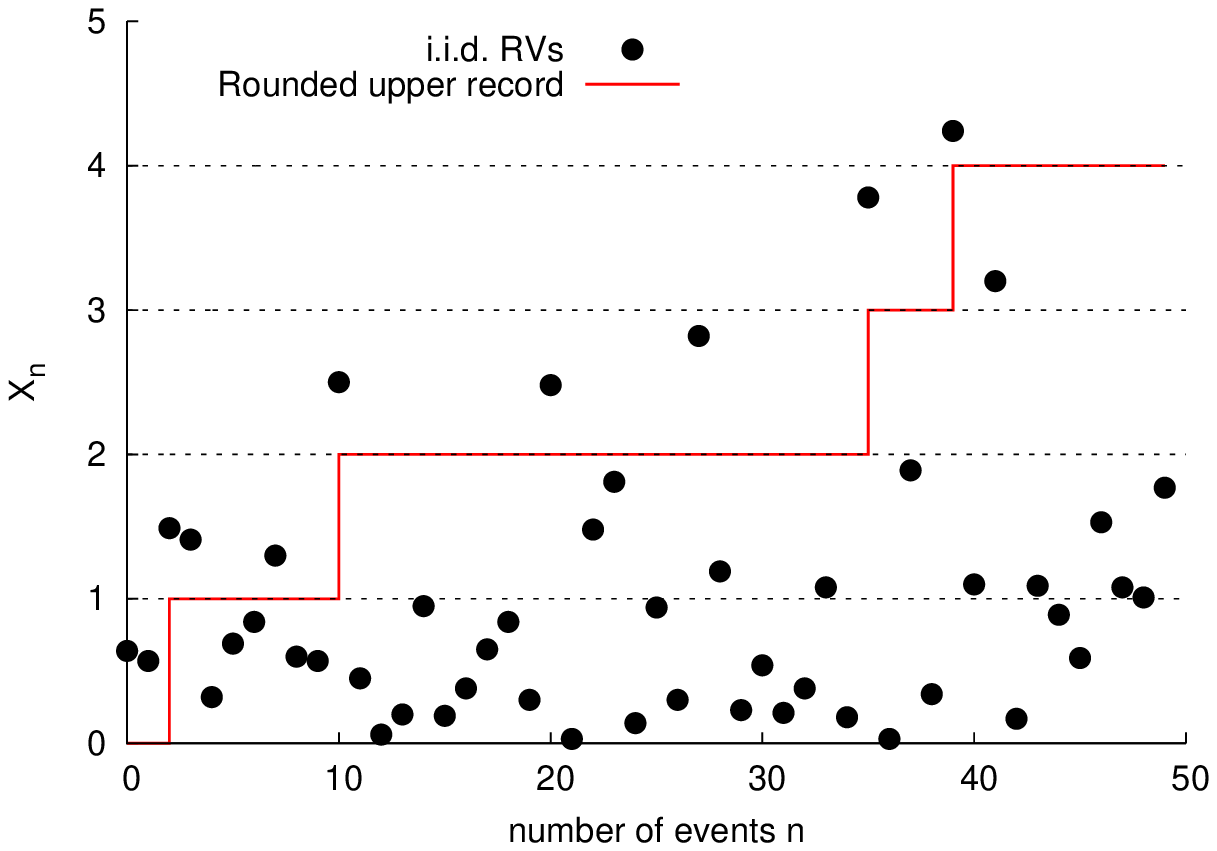}}
\centerline{\includegraphics[width=0.75\textwidth]{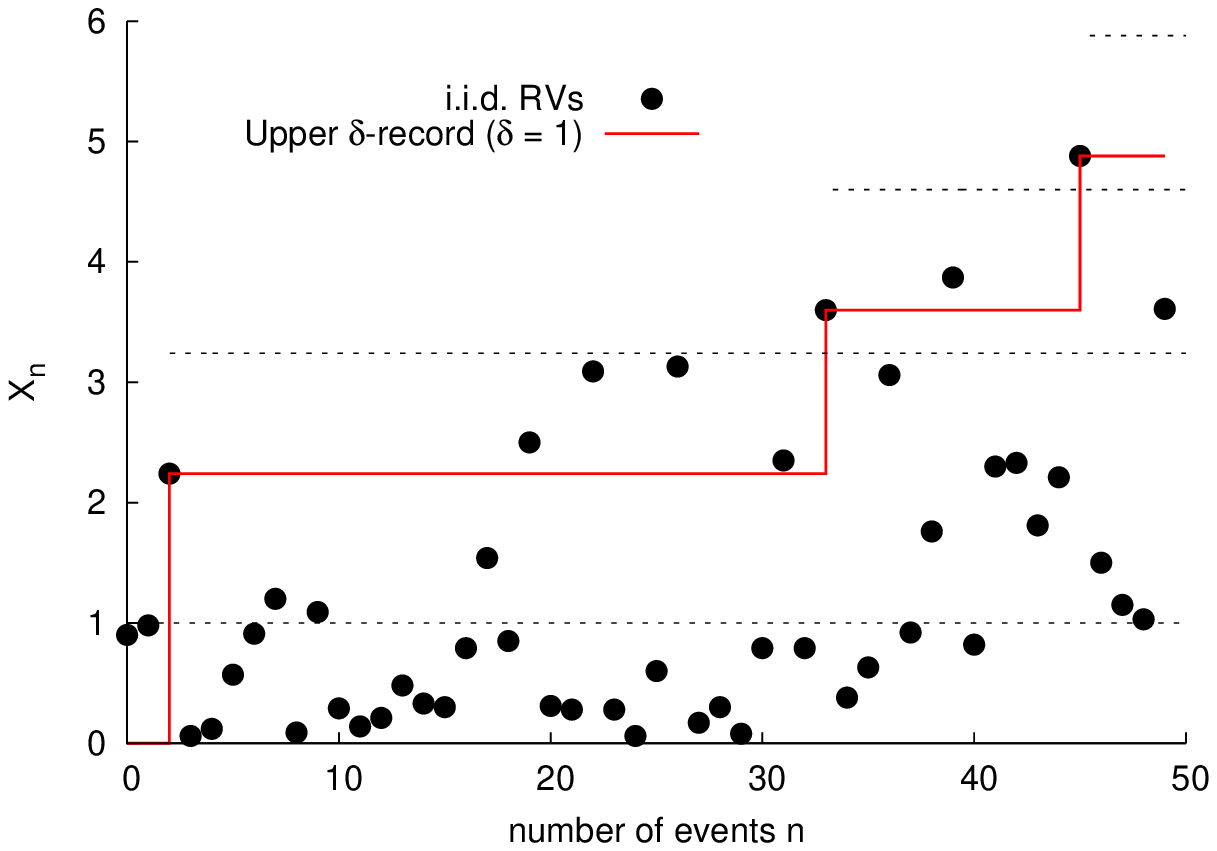}}
\caption{ \label{rev:Fig_dis} \textbf{Top: } Sketch of the record process of (exponential) i.i.d.~RV's, which are rounded down to the next integer (given by the dashed lines). A continuous RV $X_k$ (dots) is a new record if $\lfloor X_k\rfloor$ exceeds all previous values. The progression of the (upper) record value is given by the red line. \textbf{Bottom: } Sketch of the $\delta$-record process of (exponential) i.i.d.~RV's for $\delta=1$. An entry $X_k$ is counted as a new $\delta$-record, if it is larger than $\textrm{max}\{X_1,...,X_{k-1}\}+\delta$ (dashed lines). Again, the progression of the (upper) record value is given by the red line.}
\end{figure}

For the discrete uniform distribution, it is very easy to compute the strong record rate $P_n$, which is given by the following sum:
\begin{eqnarray}
 P_n = \sum_{k=1}^N \frac{1}{N}\left(\frac{k-1}{N}\right)^{n-1},
\end{eqnarray}
For $n\rightarrow\infty$ this behaves like $P_n\approx N^{-1}\left(1-N^{-1}\right)^n$, which leads, of course, to a finite mean record number $\langle R_n\rangle \propto N$. For the weak record rate $p_n$ the situation is different and one finds that the asymptotic record rate is given by $p_n\approx N^{-1}$, which leads to a divergent weak mean record number of $\langle r_n\rangle \approx n/N$.

The case of the geometric distribution is already much more complicated and was considered by Prodinger in 1996 \cite{Prodinger1996} (see also Vervaat \cite{Vervaat1973}). He derived the asymptotic mean record number $\langle R_n \rangle$ in the strong case for the geometric distribution with $P\left[X=k\right] = (1-p)^{k-1}p$. In a rather complicated, combinatorical computation, he showed that for $n\rightarrow\infty$:
\begin{eqnarray}
 \langle R_n \rangle \approx \frac{p}{\ln \left(\left(1-p\right)^{-1}\right)} \left( \ln n + \gamma - \sum_{k\neq 0} \Gamma \left(\alpha_p\right)n^{\alpha_p} \right) + \frac{p}{2}
\end{eqnarray}
With an imaginary constant $\alpha_p = \left(2k\pi i\right)/\ln\left(1-p\right)^{-1}$. The occurrence of the oscillatory term in this expression is quite surprising and, to our knowledge, it is difficult to explain this effect intuitively. 

\bigskip
Apart from that, it is also interesting to consider discreteness effects in RV's from continuous distributions. While in our considerations in sections \ref{rev:classical} and \ref{rev:uncorr} a record entry $X_n$ was simply a value that was larger than all previous values $X_0,X_1,...,X_{n-1}$, one can also impose different, more complicated, conditions, where records are only counted if they exceed another barrier depending on $X_0,X_1,...,X_{n-1}$. Some important examples that have been studied in the literature are the following:
\begin{itemize}
 \item Rounded records: An entry $X_n$ is a strong record if the rounded value $\lfloor X_n\rfloor_\Delta$ exceeds the maximum of all previous entries:
 \begin{eqnarray}
  \lfloor X_n\rfloor_\Delta > \textrm{max}\{\lfloor X_0\rfloor_\Delta,\lfloor X_1\rfloor_\Delta,...,\lfloor X_n \rfloor_\Delta\}.
 \end{eqnarray}
Here, $\lfloor \cdot\rfloor_\Delta$ means rounding (up or down) to the next integer multiple of $k\cdot \Delta$ with $k\in\mathbb{Z}$. Similarly, we have a weak record if 
\begin{eqnarray}
 \lfloor X_n\rfloor_\Delta \geq \textrm{max}\{\lfloor X_0\rfloor_\Delta,\lfloor X_1\rfloor_\Delta,...,\lfloor X_n \rfloor_\Delta\}. 
\end{eqnarray}
This process is illustrated for exponential RV's in Fig. \ref{rev:Fig_dis} (top).
 \item $\delta$-records: A $\delta$-record is an entry $X_n$ that exceeds all previous entries $X_0,X_1,...,X_{n-1}$ at least by $\delta$:
 \begin{eqnarray}
  X_n > \textrm{max}\{X_0+\delta,X_1+\delta,...,X_{n-1}+\delta\}.
 \end{eqnarray}
Note that $\delta$ can, in principle, also be negative. Such a record is called a strong record for $\delta>0$ and a weak record if $\delta<0$. This record model is sketched in Fig. \ref{rev:Fig_dis} (bottom) for $\delta=1$ and RV's from an exponential distribution.
 \item Geometric records: For a geometric record, an entry $X_n$ has to exceed a fixed multiple of all previous entries:
 \begin{eqnarray}
  X_n > \textrm{max}\{\alpha X_0,\alpha X_1,..., \alpha X_{n-1}\},
 \end{eqnarray}
where $\alpha>0$ is an arbitrary constant. Here, a record is a strong record if $\alpha>1$ and a weak record for $\alpha<1$.
\end{itemize}

In the following, we summarize how the record statistics in these three cases differ from the continuous case. As in the above, the three universality classes of EVS will play an important role. In all three cases, the findings will differ systematically between these classes.

\subsection{Rounding effects}

The record statistics of rounded measurements where first considered systematically by Wergen et al. in 2012 \cite{Wergen2012}. In a previous study, they analyzed historical temperature measurements from U.S.~weather stations \cite{USHCN} that were recorded in whole degrees of Fahrenheit. They observed that this discreteness had a significant effect on the record statistics of the temperature data that could, in principle, disguise a possible effect of global warming on the occurrence of record-breaking events \cite{Wergen2010,Wergen2012}. The problem is more general: In all applications, experimental measurements can only be recorded up to a certain accuracy. Usually, one has to deal with RV's, which are sampled from a hypothetical continuous distribution and then discretized by rounding in the measurement process. 

In 2012, Wergen et al.~\cite{Wergen2012}, studied the strong record rate and the mean record number of i.i.d.~RV's for a continuous pdf $f\left(x\right)$ that were rounded down to integer multiples of a discretization length $\Delta$. They showed that the strong record rate $P_n^{\Delta}$, in this case, can by computed from the following sum:
\begin{eqnarray}\label{rev:rec_rate_rounding}
 P_n^{\Delta} = \sum_{k} \left[ F\left(\left(k+1\right)\Delta\right) - F\left(k\Delta\right) \right] F^{n-1}\left(k\Delta\right).
\end{eqnarray}
This is just the sum over the probabilities for a new record at time $n$ on the individual lattice sites $k\Delta$ (with $k\in\mathbb{Z}$). For $\Delta\rightarrow0$, it is easy to show that $P_n^{\Delta}$ approaches the continuous result with $P_n = \int \mathrm{d}x f\left(x\right) F^{n-1}\left(x\right)$ as in section \ref{rev:classical}.

In the limit of $n\rightarrow\infty$, it is possible to analyze the asymptotic behavior of Eq. \ref{rev:rec_rate_rounding} with respect to the universality classes of EVS. For that purpose, we can again use the GPD (Eq. \ref{rev:generalized_pareto}). Since interesting results can only be expected for a discretization length $\Delta$ much smaller than the support of the distribution, we can approach the problem by replacing the sum in Eq. \ref{rev:rec_rate_rounding} by an integral. In this case, however, the bounds of integration have to be chosen carefully. Then, the strong record rate for the GPD is given by
\begin{eqnarray}\label{rev:P_nd_integrals}
 P_n^{\Delta} \approx \begin{cases} \int_1^{\frac{1}{\Delta}-1} \mathrm{d}x \left[ F\left(\left(k+1\right)\Delta\right) - F\left(k\Delta\right) \right] F^{n-1}\left(k\Delta\right), & \textrm{for } \xi < 0 \\
 \int_1^{\infty} \mathrm{d}x \left[ F\left(\left(k+1\right)\Delta\right) - F\left(k\Delta\right) \right] F^{n-1}\left(k\Delta\right), & \textrm{for }\xi \geq 0 \end{cases}
\end{eqnarray}
Here, the upper bound for the Weibull class $\frac{1}{\Delta}-1$ is simply the finite number of lattice sites in $\left[1,1-\xi^{-1}\right]$ minus one. In the case of weak records one has to omit the minus one. With Eq. \ref{rev:P_nd_integrals}, we can compute the large $n$ limit of $P_n^{\Delta}$ and find that
 \begin{eqnarray}
 P_n^{\Delta} \xrightarrow[n\rightarrow\infty]{} \begin{cases} \frac{1}{n}e^{-n\Delta^{-\frac{1}{\xi}}}, & \textrm{for } \xi < 0 \quad \textrm{(Weibull class)} \\ \frac{1}{n\Delta} \left(1-e^{-\Delta}\right) & \textrm{for } \xi = 0 \quad \textrm{(Exp. distribution)} \\ \frac{1}{n} & \textrm{for } \xi > 0 \quad \textrm{(Fr\'echet class)} \end{cases}
\end{eqnarray}
Apparently, the record rate changes systematically in the Weibull class, while, in the Fr\'echet class, the asymptotic behavior is as in the continuous case. The corresponding results for the weak record rate $p_n^{\Delta}$ are the following:
 \begin{eqnarray}
 p_n^{\Delta} \xrightarrow[n\rightarrow\infty]{} \begin{cases} \left(1+\xi\left(1-\Delta\right)\right)^{-\frac{1}{\xi}}, & \textrm{for } \xi < 0 \quad \textrm{(Weibull class)} \\ \frac{1}{n\Delta} \left(e^{\Delta}+e^{-\Delta}\right) & \textrm{for } \xi = 0 \quad \textrm{(Exp. distribution)} \\ \frac{1}{n} & \textrm{for } \xi > 0 \quad \textrm{(Fr\'echet class)} \end{cases}
\end{eqnarray}
Here, the asymptotic weak record rate in the Weibull class is constant and equals the probability that a RV falls into the largest lattice site. Both cases show that rounding effects are important for RV's from the Weibull class, while they are negligible in Fr\'echet class, where, because of the heavy tails of the distributions, the asymptotic record rate always converges to the $1/n$-behavior of the record process without rounding. The behavior in the Gumbel class is, as usual, intermediate between those two. While the (strong and weak) record rates of the exponential distribution are still proportional to $1/n$, Wergen et al.~\cite{Wergen2012} found sublinear corrections to the $1/n$-behavior for the GGD with $f\left(x\right) \propto e^{-|x|^{\beta}}$. Here, for $\beta>1$, the strong and weak record rates decay as
\begin{eqnarray}
 P_n^{\Delta} \propto \frac{1}{n} \ln n^{\frac{1}{\beta}-1}\quad\quad\quad \textrm{and} \quad\quad\quad p_n^{\Delta} \propto \frac{1}{n} \ln n^{1-\frac{1}{\beta}}.
\end{eqnarray}
Note that, even though these results were derived for the special case of 'rounding down', they do not change systematically if one considers other kinds of rounding like 'rounding up' or 'rounding to the nearest integer'.

Wergen et al.~\cite{Wergen2012} also considered the interesting regime of very strong discreteness with $\Delta\gg 1$. Here, the occurrence of records becomes predictable on a logarithmic time-scale for certain distributions from the Gumbel class. For a detailed discussion of this phenomenon we refer the reader to \cite{Wergen2012}.

\subsection{$\delta$-records}

The concept of $\delta$-records (or near-records) was discussed by various authors, for instance by Gouet et al.~\cite{Gouet2005,Gouet2007,Gouet2012} or Balakrishnan et al.~\cite{Balakrishnan1996,Balakrishnan2005}. In particular Gouet et al. made important progress on this problem. In \cite{Gouet2007}, they discussed the process of $\delta$-records in detail and, using a so-called Martingale approach (see \cite{Gouet2007} and references therein), proved a limit theorem for the asymptotic distribution of the record number in this case. Instead of describing their rather complex derivations, we will now demonstrate an elementary approach that illustrates the asymptotic behavior of $\delta$-records in time series of RV's from the three classes of EVS in the regime of small $\delta\ll1$. Our findings are in good agreement with the results of Gouet et al.~\cite{Gouet2007}.

In the general case, it is easy to see that the record rate of $\delta$-records can be obtained from the integral
\begin{eqnarray}
 P_n^{\delta} = \int \mathrm{d}x\;f\left(x\right) F^{n-1}\left(x-\delta\right).
\end{eqnarray}
Again, for $\delta=0$, we obtain the continuous result with $P_n=1/n$. As in the case of the LDM (see section \ref{rev:uncorr}) we can now expand this integral for small values of $\delta$ and $n\gg1$. Doing this we find
\begin{eqnarray}
 P_n^{\delta} & \approx & \int \mathrm{d}{x}\; f\left(x\right)\left(F^{n-1}\left(x\right) - \delta \left(n-1\right) f\left(x\right) F^{n-2}\left(x\right)\right) \nonumber \\
 & \approx & \frac{1}{n+1} - \delta n \textrm{I}_n.
\end{eqnarray}
with the same $\textrm{I}_n = \int \mathrm{d}x\;f^2\left(x\right) F^{n-2}\left(x\right)$ as in section \ref{rev:uncorr}. With the results for $\textrm{I}_n$ described in that section, we can now compute the rate of $\delta$-records for RV's from a GPD in the regime of small $\delta$. Here, we find that
\begin{eqnarray}
 P_n^{\delta} \approx \frac{1}{n+1} - \delta \begin{cases} 2\xi\Gamma\left[2+\xi\right] n^{-\xi-1}, & \textrm{for } \xi \neq 0 \\ \frac{1}{n+1}, & \textrm{for } \xi = 0                                          
  \end{cases}
\end{eqnarray}
For $\xi\geq 0$ (Exponential distribution and Fr\'echet class) this result is correct even in the limit $n\rightarrow\infty$. In the Weibull class with $\xi<0$ it holds for $\delta\ll n^{\xi}$. The result illustrates nicely how the $\delta$ affects the record rate in the $\delta\ll 1$ regime. As in the case of rounding discussed before, the $\delta$ is negligible in the Fr\'echet class and has a strong effect that increases with $n$ in the Weibull class. It is straightforward to show that, in the Weibull class, the record rate will eventually decay exponentially, which leads to a finite asymptotic record number \cite{Gouet2005,Gouet2007}.

Again, the case of the Gumbel class is more complicated. For the GGD with $f\left(x\right) \propto e^{-|x|^{\beta}}$ one finds a that, for small $\delta \ll 1$
\begin{eqnarray}
 P_n^{\delta} \approx \frac{1}{n+1} - A_{\beta} \frac{\delta}{n} \ln \left(n\right)^{1-\frac{1}{\beta}}. 
\end{eqnarray}
With a positive constant $A_{\beta}$, which depends on the tail parameter $\beta$. While, for $\beta<1$, this approximation is valid for arbitrary values of $n$, it only holds for $\ln\left(n\right)^{1-\frac{1}{\beta}} \ll \delta^{-1}$ when $\beta$ is larger than one. This result indicates that the marginal case of $\beta=1$ plays an important role. 

In fact, it is well known that the mean spacings $\langle\Delta_k\rangle$ between the subsequent records with record numbers $k$ and $k+1$ from an exponential distribution are equidistant from each other \cite{Arnold1998}. For all (light-tailed) distributions decaying faster than the exponential, e.g.~with $\beta>1$, these mean spacings are decreasing with increasing $k$ and the record values move closer and closer together. For $\beta<1$ and (heavy-tailed) distributions broader than the exponential, the spacings increase with $k$. Only in the regime of $\beta>1$, the spacings will eventually become smaller than any $\delta$. Eventually, as shown rigorously by Gouet et al.~\cite{Gouet2012}, for very large $n$, this leads to a slow exponential decay of the record rate for all distributions with $\beta>1$. Using the results of Gouet et al.~\cite{Gouet2012} one can compute the (exact) asymptotic mean record number for the GGD in the large $n$ limit:
\begin{eqnarray}
 \langle R_n\rangle \xrightarrow[n\rightarrow\infty]{} \begin{cases} \frac{B_{\beta}}{\delta \left(\beta-1\right)} \ln \left(n\right)^{1-\frac{1}{\beta}} e^{-\delta C_{\beta} \ln \left(n\right)^{\beta+\frac{1}{\beta}-2}}, & \textrm{for } \beta > 1 \\
 \ln \left(n\right) e^{-\delta}, & \textrm{for } \beta = 1 \\ 
 \ln \left(n\right), & \textrm{for } \beta < 1 \end{cases}
\end{eqnarray}
with positive constants $B_{\beta}$ and $C_{\beta}$ depending on $\beta$.

\subsection{Geometric records}

The first author who discussed the problem of geometric records was Eliazar in 2005 \cite{Eliazar2005}. A geometric record is a record that is only counted if it exceeds a certain multiple of the previous record. In particular, in order to be a record, $X_n$ has to be larger than $\alpha\cdot\textrm{max}\{X_0,X_1...,X_n\}$ with a positive constant $\alpha$ (not necessarily larger than one). The record rate $P_n^{\alpha}$ for this problem is given by
\begin{eqnarray}\label{rev:rec_rate_geometric}
 P_n^{\alpha} = \int \mathrm{d}x\; f\left(x\right) F\left(\frac{x}{\alpha}\right)^{n-1}.
\end{eqnarray}
For the exponential distribution with $f\left(x\right)$ ($x>0$), this integral can be computed exactly. Here we find
\begin{eqnarray}
 P_n^{\alpha} = \int_0^{\infty} \mathrm{d}x\;e^{-x}\left(1-e^{-\frac{x}{\alpha}}\right)^{n-1} \; \xrightarrow[n\rightarrow\infty]{} \; \frac{\alpha \Gamma\left[\alpha\right]}{n^{\alpha}},
\end{eqnarray}
which reproduces the i.i.d.~record rate for $\alpha=1$. Interestingly, for $\alpha>1$, this indicates a finite asymptotic mean record number. For the GPD with $\xi\neq 0$, the situation is more complicated and we can not compute the record rate $P_n^{\alpha}$ directly for other representatives of the distribution. In a recent article, Gouet et al.~\cite{Gouet2012} proved a series of theorems for the record rate of geometric records that can be used to give the asymptotic record rate of the GPD in the geometric case. 

Gouet et al. showed that the record rate $P_n^{\alpha}$ of distributions of the Fr\'echet class does not differ significantly from the i.i.d.~case. In the Weibull class, on the other hand, the asymptotic record rate goes to zero for $\alpha>1$. In the Gumbel class, the situation is more complicated and, for $\alpha>1$, the mean record number can be both finite or divergent. Interestingly, in this class one also finds distributions where the mean record number goes to infinity with a slower than logarithmic speed. With the results presented in \cite{Gouet2012}, we can infer the record rate $P_n^{\alpha}$ for the GPD for $n\rightarrow\infty$ and $\xi\geq 0$:
\begin{eqnarray}
 P_n^{\alpha} \;\xrightarrow[n\rightarrow\infty]{} \begin{cases} 
\alpha \Gamma\left[\alpha\right]n^{-\alpha}, & \textrm{for } \xi = 0 \quad \textrm{(Exp. distribution)} \\
\alpha^{-\frac{1}{\xi}} n^{-1}, & \textrm{for } \xi > 0 \quad \textrm{(Fr\'echet class)} 
      \end{cases}
\end{eqnarray} 
In the Weibull class, we expect an exponential decay of $P_n^{\alpha}$, but, so far, we are not aware of any analytical results for this regime.

\section{Records in correlated processes}
\label{rev:corr}

\subsection{Records in symmetric, discrete-time random walks}
\label{rev:symmetric_rw}

\begin{figure}[t]
\centerline{\includegraphics[width=0.75\textwidth]{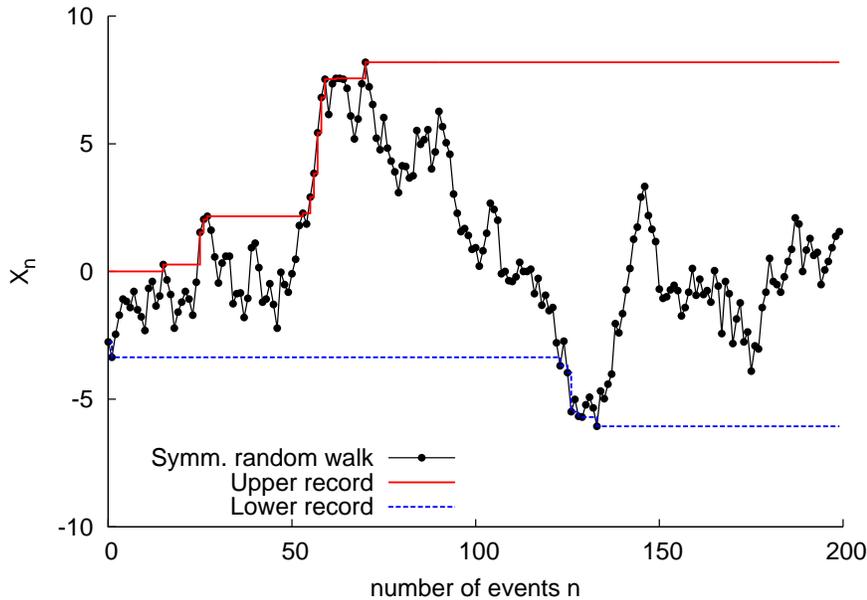}}
\caption{\label{rev:Fig_RW} Sketch of a symmetric random walk with a Gaussian jump distribution. The red (blue dotted) lines mark the progression of the upper (lower) record of the process.}
\end{figure}

An entirely new field of research was established through the work of Majumdar and Ziff \cite{Majumdar2008}, who were the first to consider the record statistics of symmetric random walks. In contrast to most of the previous research in the field of record statistics, they considered a correlated process, namely a symmetric, discrete-time random walk (an introduction can be found in \cite{Weiss1994,Feller1968}), and computed its record rate, its mean record number and also the full distribution of the record number. In the following, we will summarize and discuss their important findings. 

A discrete-time random walk (DTRW) $X_0,X_1,...,X_n$ is a time series with entries of the form
\begin{eqnarray}\label{rev:X_i_symm_RW}
 X_i = X_{i-1} + \eta_i,
\end{eqnarray}
with i.i.d.~increments $\eta_i$ drawn from a continuous and symmetric distribution $f\left(\eta\right)$ (see also Fig. \ref{rev:Fig_RW}). Without loss of generality, we can set $X_0=0$. Then, by definition, $X_0=0$ is also the first record.

To compute the record statistics of this process, it is helpful to introduce two generally important quantities, the first-passage probability $\phi\left(x,n\right)$ and the survival probability $q\left(x,n\right)$ (cf. \cite{RednerBook}). The (positive) first-passage probability is the probability that a random walk, starting at $0$, crosses $x\geq0$ in time-step $n$ for the first time:
\begin{eqnarray}
 \phi\left(x,n\right) := \textrm{Prob}\left[X_n>x\;\&\;X_0,X_1,...,X_{n-1}\leq x\right].
\end{eqnarray}
The related (positive) survival probability $q\left(x,n\right)$ is the probability that the random walk remains below $x$ for the first $n$ steps:
\begin{eqnarray}
 q\left(x,n\right) := \textrm{Prob}\left[X_0,X_1,...,X_n\leq x\right].
\end{eqnarray}
It is easy to see that the first-passage probability can also be obtained by $\phi\left(x,n\right) = q\left(x,n-1\right) - q\left(x,n\right)$. 

In the special case of a symmetric DTRW ($f\left(\eta\right) = f\left(-\eta\right)$) and $x=0$, these quantities can be computed using an important theorem by Sparre Andersen \cite{SparreAndersen1953,SparreAndersen1954}. He showed that, in this case, the generating function of the survival probability $q\left(0,n\right)$, defined as $\tilde{q}\left(0,z\right) = \sum_{n=0}^{\infty} q\left(0,n\right) z^n$ is given by
\begin{eqnarray}\label{rev:sparre_andersen}
 \tilde{q}\left(0,z\right) = \frac{1}{\sqrt{1-z}}.
\end{eqnarray}
Expanding in powers of $z$ this leads to $q\left(0,n\right) = \left( 2n \atop n \right) 2^{-2n}$. 

This result was the most important requirement for the work of Majumdar and Ziff \cite{Majumdar2008}. They showed that a random walk of length $n$ with $R_n$ records can be described as a chain of $R_n - 1$ first-passage problems and one survival problem. This is possible because of the so-called \textit{renewal property} of the random walk. After a record at time $i$, the probability for a record at time $i+j$ is the same as the probability $\phi\left(0,j\right)$ that a random walk starting from $0$ crosses the origin (from negative to positive) after $j$ steps for the first time. As long as the process stays below the origin set by the record at time $i$, no further records occur.

Therefore, the probability $P\left(i_1,...,i_{R_n};n\right)$ for a random walk with records at times $0,i_2,i_3,...,i_{R_n}$ (with $i_1=0$ by definition and $0<i_2<i_3<...<i_{R_n}\leq n$) can be given by
\begin{eqnarray}
P\left(i_1,...,i_{R_n};n\right) = \quad\quad\quad\quad\quad\quad\quad\quad\quad\quad\quad\quad\quad\quad\quad\quad\quad\quad\quad\quad\quad\quad \nonumber \\ \phi\left(0,i_2\right)\cdot \phi\left(0,i_3-i_2\right)\cdot ... \cdot \phi\left(0,i_{R_{n}}-i_{R_{n-1}}\right)\cdot q\left(0, n-i_{R_{n}}\right) 
\end{eqnarray}
With this, the distribution $P\left(R_n|n \right)$ of the record number $R_n$ can be obtained by summing over all possible sets of inter-record times $0,i_2,i_3,...,i_{R_n}$ with $0<i_2<...<i_{R_n}\leq n$. The easiest way to compute this sum is via the generating function of $P\left(R_n|n \right)$. Majumdar and Ziff found that $P\left(R_n|n \right)$ obeys 
\begin{eqnarray}\label{rev:gen_function_prn_n}
 \sum_{n=R_n-1}^{\infty} P\left(R_n|n\right) z^n & = & \left(\tilde{\phi}\left(z\right)\right)^{R_n-1} \tilde{q}\left(z\right) \nonumber \\
 & = & \left(1-\left(1-z\right)\tilde{q}\left(z\right)\right)^{R_n-1} \tilde{q}\left(z\right) 
\end{eqnarray}
and, with the survival probability $\tilde{q}\left(z\right) = \sqrt{1-z}^{-1}$ of the symmetric random walk, one finds
\begin{eqnarray}
\sum_{n=R_n-1}^{\infty} P\left(R_n|n\right) z^n = \frac{\left(1-\sqrt{1-z}\right)^{R_n-1}}{\sqrt{1-z}}.
\end{eqnarray}
This result allowed Majumdar and Ziff to extract the exact distribution of the record number $R_n$:
\begin{eqnarray}
 P\left(R_n|n \right) = \left( 2n - R_n + 1 \atop n \right) 2^{-2n + R_n - 1}.
\end{eqnarray}
From this expression, one can easily obtain the mean record number $\langle R_n \rangle$ and the record rate $P_n$ of the symmetric DTRW. For the generating function of $\langle R_n \rangle$, one has to multiply Eq. \ref{rev:gen_function_prn_n} with the record number $R_n$ and sum over all possible values for $R_n$. This leads to
\begin{eqnarray}
 \sum_{n=0}^{\infty} \langle R_n \rangle z^n & = & \sum_{R_n=0}^{\infty} R_n \left(\tilde{\phi}\left(z\right)\right)^{R_n-1}  \tilde{q}\left(z\right) = \frac{1}{\left(1-z\right)^{3/2}}. 
\end{eqnarray}
Expanding this result in powers of $z$ we find
\begin{eqnarray}
 \langle R_n \rangle = \left(2n+1\right)\left( 2n \atop n\right) 2^{-2n} \quad\quad\quad \textrm{and} \quad\quad\quad P_n = \left(2n \atop n\right) 2^{-2n}.
\end{eqnarray}
Is is interesting to analyze the asymptotic behavior of these quantities in the limit of $n\rightarrow\infty$. Here, the record number approaches a half-Gaussian distribution with 
\begin{eqnarray}
 P\left(R_n|n \right) \approx \frac{1}{\sqrt{n\pi}} e^{-\frac{R_n^2}{4n}}.
\end{eqnarray}
The mean record number and the record rate converge to
\begin{eqnarray}
 \langle R_n \rangle \approx \sqrt{\frac{4n}{\pi}} \quad\quad\quad \textrm{and} \quad\quad\quad P_n \approx \frac{1}{\sqrt{\pi n}}.
\end{eqnarray}
Majumdar and Ziff also considered discrete random walks on a lattice with lattice constant $d$ and a jump distribution $f\left(x\right) = \frac{1}{2}\left(\delta\left(x-d\right) + \delta\left(x+d\right)\right)$. In this case, the asymptotic record statistics is very similar to the continuous case and differs only in a prefactor. For $n\rightarrow\infty$, the mean record number and the record rate are reduced by a factor of $1/\sqrt{2}$:
\begin{eqnarray}
 \langle R_n \rangle \approx \sqrt{\frac{2n}{\pi}} \quad\quad\quad \textrm{and} \quad\quad\quad P_n \approx \frac{1}{\sqrt{2\pi n}}.
\end{eqnarray}
In the article by Majumdar and Ziff \cite{Majumdar2008} one can also find a discussion of the extreme value statistics of the ages of the longest and shortest lasting records in a symmetric random walk. In particular, they showed that the expected age of the longest lasting record grows proportional to the walk length $n$ and not to $\sqrt{n}$ as the average age of a record and also the age of the shortest lasting record.

\subsection{Biased random walks}
\label{rev:sec_biased_rw}

A natural way to generalize the model of a symmetric DTRW considered by Majumdar and Ziff, is to introduce a bias. The entries $X_0,X_1,..,X_n$ of such a biased random walk with a constant drift $c$ are given by
\begin{eqnarray}
 X_i = X_{i-1} + \eta_i + c,
\end{eqnarray}
where the $\eta_i$'s are i.i.d.~RV's from a symmetric distribution $f\left(\eta\right)$ as in the previous section and again $X_0=0$. As in the case of the LDM for uncorrelated time series, the drift causes a non-universal and distribution dependent behavior of the record statistics of the DTRW. The simple, universal version of the Sparre Andersen theorem is not valid in the biased case and the first-passage and survival probabilities of the random walk depend on the choice of the jump distribution $f\left(\eta\right)$.

Fortunately, there exists a more general version of Sparre Andersen's theorem that holds also in the biased case. Sparre Andersen \cite{SparreAndersen1953,SparreAndersen1954} showed that the (positive) survival probability with respect to the origin of the biased random walk $q_c\left(0,n\right)$ has the generating function
\begin{eqnarray}\label{rev:gen_sparre_andersen}
\tilde{q}\left(0,z\right) = \textrm{exp}\left(\sum_{n=1}^{\infty} \frac{z^n}{n!} \rho_c\left(n\right)\right),
\end{eqnarray}
where $\rho_c\left(n\right)$ is the probability that a random walk is negative at time $n$: $\rho_c\left(n\right):=P\left[X_n<0\right]$. In the case of an unbiased random walk with $c=0$, we have $\rho_0\left(n\right) = \frac{1}{2}$, which reduces Eq. \ref{rev:gen_sparre_andersen} to the simple symmetric version of the Sparre Andersen theorem introduced in the previous section (Eq. \ref{rev:sparre_andersen}). In the general case, it is more complicated to compute $\rho_c\left(n\right)$ and the probability density of the random walk at time $n$ is needed.

Majumdar et al.~\cite{Majumdar2012} found that the asymptotic behavior of $q_c\left(0,n\right)$ depends crucially on the tail of the jump distribution $f\left(\eta\right)$. Since, the behavior of the distribution $f\left(\eta\right)$ for large values of $\eta$ is dictated by the small $k$ behavior of its Fourier transform $\tilde{f}\left(k\right)$, Majumdar et al. considered jump distributions with Fourier representations of the form
\begin{eqnarray}\label{rev:f_k}
 \hat{f}\left(k\right) \approx 1 + \left(l_{\mu}|k|\right)^{\mu}
\end{eqnarray}
for small values of $k$. Here, $l_{\mu}$ is a constant parameter and $\mu \in \left(0,2\right]$ the so-called L\'evy-index (also the tail-index) of the jump distribution. While a L\'evy-index with $\mu=2$ corresponds to jump-distributions with a finite second moment $\sigma^2 = \int \mathrm{d}x\; x^2 f\left(x\right)$, a value of $\mu<2$ describes a distribution with heavy-tails, whose variance does not converge. In real-space, the tails of such a distribution decay as $f\left(x\right) \propto |x|^{-\mu-1}$ for $x\rightarrow\infty$. Distributions with $\mu\leq1$ have even a divergent mean value $\int \mathrm{d}x\;xf\left(x\right)$.

For the simple form of $\tilde{f}\left(k\right)$ in Eq. \ref{rev:f_k} one can compute $\rho_c\left(n\right)$ by elementary means. Obtaining the generating function $\tilde{q}\left(0,z\right)$ in a closed form is, however, far more complicated and requires more sophisticated techniques.

Without going into the details of the computations of Majumdar et al. \cite{Majumdar2012}, we will now summarize their results for the survival probability $q_c\left(0,n\right)$, the distribution of the record number $P\left(R_n|n\right)$, the mean record number $\langle R_n\rangle$ and the record rate $P_n$ of the biased random walk. In fact, they found five different universal regimes depending on the bias $c$ and the L\'evy-index $\mu$, in which the asymptotic survival and record statistics have systematically different characteristics. These regimes are the following:

\paragraph{I --- The subcritical case ($\mu<1$):}

In this regime, the standard deviation of the position of the random walker $X_n$ grows faster than linear and the effects of the drift are therefore negligible in the large $n$ limit. In fact, the survival probability $q_c\left(0,n\right)$ is proportional to $1/\sqrt{n}$ as in the unbiased case. The mean record number and the distribution of the record number have the same large $n$ behavior as the symmetric random walk, only their prefactors are different. Also the extremal ages of the shortest and longest lasting records have the same large $n$ asymptotics as in the symmetric case.

\paragraph{II --- The marginal case ($\mu=1$):}

In this interesting regime, the survival and record statistics have a more complicated dependence on $n$. In some sense, this is the regime, were the drift and the fluctuations of the process are of the same order. Here, the survival probability $q_c\left(0,n\right)$ decays like 
\begin{eqnarray}
 q_c\left(0,n\right) \propto \frac{1}{n^{\Theta\left(c\right)}},
\end{eqnarray}
with a $c$ dependent exponent $\Theta\left(c\right) = \frac{1}{2} + \frac{1}{\pi} \textrm{arctan}\left(c\right)$. This non-trivial exponent also appears in the mean record number. Majumdar et al. \cite{Majumdar2012} showed that
\begin{eqnarray}
 \langle R_n\rangle \propto n^{\Theta\left(c\right)} \quad\quad\quad \textrm{and} \quad\quad\quad P_n \propto n^{\Theta\left(c\right) - 1}.
\end{eqnarray}
Also the full distribution of the record number $P\left(R_n|n\right)$ and the extremal ages of the shortest and longest lasting records depend on this function $\Theta\left(c\right)$. In both cases the asymptotic results are more complicated and we refer to \cite{Majumdar2012} for details.

A jump distribution, which falls into this regime, is the Cauchy distribution with $f\left(x\right) = \frac{1}{\pi\left(1+x^2\right)}$. This special case was already considered by Le Doussal and Wiese \cite{LeDoussal2009}, prior to the work of Majumdar et al.. They also found the non-trivial exponent $\Theta\left(c\right)$ and computed the exact mean record number, as well as its variance, for this case. For a biased random walk with a Cauchy jump distribution the mean record number reads
\begin{eqnarray}
 \langle R_n \rangle = \frac{\Gamma\left[n+2-\Theta\left(c\right)\right]}{\Gamma\left[n+1\right]\Gamma\left[2-\Theta\left(c\right)\right]}.
\end{eqnarray}
Interestingly, the function $\Theta\left(x\right)$ is also the cumulative distribution $\int^x\mathrm{d}x\; f\left(x\right)$ of the Cauchy distribution. The reason for this agreement is, to our knowledge, unclear.

\paragraph{III --- The supercritical case with positive drift ($\mu>1$ \& $c>0$):}
Here, the survival probability decays faster than in the two previous cases. For $n\rightarrow\infty$, $q_c\left(0,n\right)$ behaves like
\begin{eqnarray}
 q_c\left(0,n\right) \propto \frac{1}{n^{\mu}}
\end{eqnarray}
and the mean record number grows linearly with $n$:
\begin{eqnarray}
 \langle R_n \rangle \approx \alpha_{\mu}\left(c\right) n
\end{eqnarray}
with a parameter $\alpha_{\mu}\left(c\right)$, which was also computed by Majumdar et al. \cite{Majumdar2012}. The distribution of the record number has an interesting, non-trivial scaling form. In particular, $P\left(R_n,n\right)$ is given by
\begin{eqnarray}
 P\left(R_n,n\right) \xrightarrow[n\rightarrow\infty]{} \frac{1}{\alpha_{\mu}\left(c\right) n^{\frac{1}{\mu}}} V_{\mu}\left(\frac{R_n - \alpha_{\mu}\left(c\right) n}{\alpha_{\mu}\left(c\right)n^{\frac{1}{\mu}}}\right)
\end{eqnarray}
The scaling function $V_{\mu}\left(u\right)$ is of the form 
\begin{eqnarray}
 V_{\mu}\left(u\right) \approx c_1 u^{\frac{2-\mu}{2\left(\mu-1\right)}} e^{-c_2 u^{\frac{\mu}{\mu-1}}}
\end{eqnarray}
with constants $c_1$ and $c_2$, which depend only on $\mu$. Here, the age of the longest lasting record grows like the inverse survival probability $\propto n^{-\mu}$ and the age of the shortest approaches a constant value proportional to $1-\alpha_{\mu}\left(c\right)$.

\paragraph{IV --- The Brownian case with positive drift ($\mu=2$ \& $c>0$):}

This is the regime of a Brownian random walk with a jump distribution that has a finite variance. Here, a positive drift has a strong effect on the survival probability and the mean record number. In fact, in contrast to regime III, the survival probability decays exponentially. Majumdar et al. \cite{Majumdar2012} showed that
\begin{eqnarray}
 q_c\left(0,n\right) \propto \frac{1}{n^{\frac{3}{2}}} e^{-\frac{c^2 n}{2\sigma^2}},
\end{eqnarray}
where $\sigma$ is the standard deviation of the jump distribution ($\sigma^2 = \int \mathrm{d}x\; x^2 f\left(x\right)$) . Despite of this systematic difference between the survival probabilities of regime III and IV, the mean record number has the same behavior. As in regime III, we have $\langle R_n \rangle \approx \alpha_2\left(c\right) n$ (with a distribution dependent prefactor $\alpha_2\left(c\right) = \lim_{\mu\rightarrow2} \alpha_{\mu}\left(c\right)$) and an asymptotically constant record rate $P_n = \alpha_2\left(c\right) > 0$. Again in contrast to regime III, the distribution of the record number is Gaussian with mean value $\langle R_n\rangle$ and a standard deviation, which grows proportional to $\sqrt{n}$. The age of the longest lasting record grows logarithmically $\propto \ln n$, while the age of the shortest approaches a constant value as in regime III.

\paragraph{V --- The supercritical case with negative drift ($\mu>1$ \& $c<0$):}

The regime with $\mu>1$ and negative drift is characterized by an asymptotically constant survival probability as well as a finite record number. Here, the drift eventually dominates the behavior and, beyond a certain time, records will no longer be possible. Majumdar et al.~\cite{Majumdar2012} computed the asymptotic survival probability $q_c\left(0,n\right) \approx a_{\mu}\left(c\right)$ and showed that the asymptotically finite mean record number is given by the inverse of this value: 
\begin{eqnarray}
 \langle R_n\rangle \approx \frac{1}{q_c\left(0,n\right)} \approx \frac{1}{a_{\mu}\left(c\right)}.
\end{eqnarray}
Here, the parameter $a_{\mu}\left(c\right)$ for $c<0$ is related to the parameter $\alpha_{\mu}\left(c\right)$ for $c>0$ from the regimes III and IV. One finds that $a_{\mu}\left(c\right) = \alpha_{\mu}\left(|c|\right)$. The distribution of the record number for $n\rightarrow\infty$ has a simple geometric form:
\begin{eqnarray}
 P\left(R_n|n\right) \approx a_{\mu}\left(c\right) \left(1-a_{\mu}\left(c\right)\right)^{R_n - 1}.
\end{eqnarray}
Due to the fact that the record number is finite, the ages of the shortest and longest lasting records grow linearly with $n$ in regime V.

\bigskip
In 2011, Wergen et al.~\cite{Wergen2011b} also considered the Brownian case (regime IV), but focused on the behavior of the record statistics for finite $n$ in the regime of a small drift $c$. Wergen et al. showed that, in this case, the survival probability and the record number of any biased random walk with a jump distribution that has a finite variance $\sigma^2$ ($\mu=2$) is very similar to the corresponding quantities of a Gaussian random walk with the same $\sigma^2$. Expanding the survival probability $q_c\left(0,n\right)$ from Eq. \ref{rev:gen_sparre_andersen} up to first order in $c$, one finds that, for $c\sqrt{n}\ll\sigma$,
\begin{eqnarray}
 q_c\left(0,n\right) \approx \frac{1}{\sqrt{\pi n}} + \frac{c}{\sqrt{2}{\sigma}}.
\end{eqnarray}
With the methods described in section \ref{rev:symmetric_rw} this result can be used to obtain the mean record number and the record rate in the regime of $c\sqrt{n}\ll\sigma$:
\begin{eqnarray}
 \langle R_n\rangle & \approx & \sqrt{\frac{4n}{\pi}} + \frac{c}{\sigma}\frac{\sqrt{2}}{\pi}\left(n\arctan\left(\sqrt{n}\right)\right), \\ P_n & \approx & \frac{1}{\sqrt{\pi n}} + \frac{c}{\sigma}\frac{\sqrt{2}}{\pi} \arctan\left(\sqrt{n}\right)
\end{eqnarray}
For $n\gg1$, the arctan-function approaches $\frac{\pi}{2}$, which leads to $\langle R_n\rangle \approx \sqrt{\frac{4n}{\pi}} + \frac{cn}{\sqrt{2}\sigma}$ and $P_n\approx \frac{1}{\sqrt{\pi n}} + \frac{c}{\sqrt{2}\sigma}$.

\subsection{Multiple random walks}
\label{rev:sec_multiple_random_walks}

\begin{figure}[t]
\centerline{\includegraphics[width=0.75\textwidth]{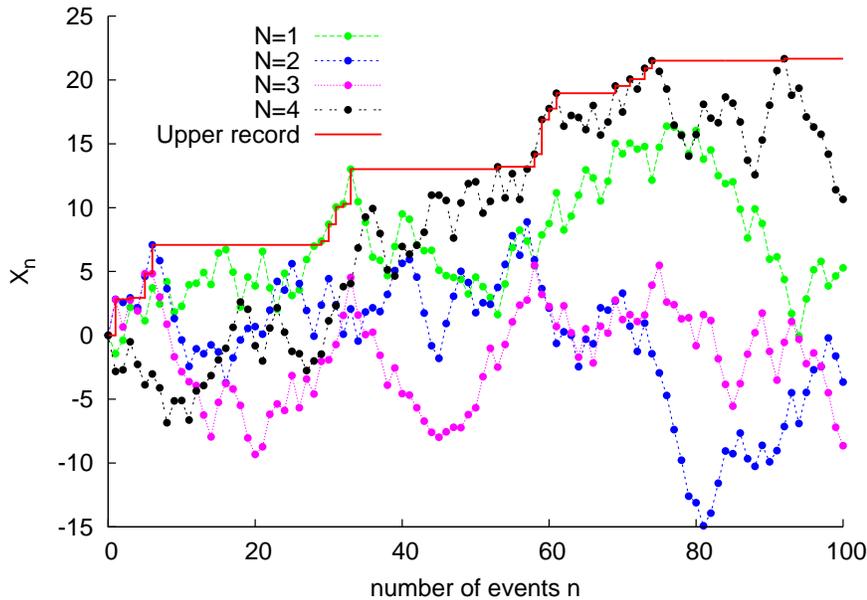}}
\caption{\label{rev:Fig_NRWs} Sketch of the record process of the maximum $X_{\textrm{max},n}\left(N\right)$ of $N=4$ independent (Gaussian) random walks. The progression of the upper record of $X_{\textrm{max},n}\left(N\right)$ is indicated by the red line. }
\end{figure}

Another way to generalize the fundamental work of Majumdar and Ziff \cite{Majumdar2008} is to consider ensembles of multiple random walks. In 2012, Wergen et al.~\cite{Wergen2012} discussed the record statistics of the maximum $X_{\textrm{max},n}\left(N\right)$ of $N$ uncorrelated DTRW's with a symmetric jump distribution. For a $N$ random walks with entries
\begin{eqnarray}
 X_{i,n} = X_{i,n-1} + \eta_{i,n} \quad\quad\quad \left(i=1,...,N\textrm{ and }X_{i,0} = 0\right) 
\end{eqnarray}
and jumps $\eta_{i,n}$ drawn from a single symmetric jump distribution $f\left(\eta\right)$. The maximum of $N$ random walks $X_{\textrm{max},n}\left(N\right)$ is defined as
\begin{eqnarray}
 X_{\textrm{max},n}\left(N\right) = \textrm{max}\{X_{1,n},X_{2,n},...,X_{N,n}\}.
\end{eqnarray}
Fig. \ref{rev:Fig_NRWs} illustrates the record process of $X_{\textrm{max},n}\left(N\right)$ for $N=4$ independent random walks. 

Unfortunately, since the maximum of $N$ random walks does not exhibit the same renewal property as the single random walk, it is impossible to compute the record statistics from the survival probability $q\left(0,n\right)$ and Sparre Andersen's theorem \cite{SparreAndersen1953,SparreAndersen1954} is not useful here.

Because of that, it is not possible to compute the probability $P\left(R_n\left(N\right),n\right)$ for $R_n\left(N\right)$ records of the maximum $X_{\textrm{max},n}\left(N\right)$ of the $N$ random walks directly. However, Wergen et al.~\cite{Wergen2012} found a more general way to calculate the record rate $P_n\left(N\right)$ that also works in absence of the renewal property.

Wergen et al.~\cite{Wergen2012} argued that the probability that the maximum of $N$ random walks sets a new record with record value $x$ at time $n$, is given by $N$ times the first-passage probability $\phi\left(x,n\right)$ multiplied with the survival probability $q\left(x,n\right)$ to the power $N-1$. This is because, $N\phi\left(x,n\right)q\left(x,n\right)^{N-1}$ is just the probability that the value $x$ is first exceeded by one of the $N$ walkers in step $n$, while the other $N-1$ stay below $x$. Integration over all possible record values $x$ leads to
\begin{eqnarray}\label{rev:P_n_N}
 P_n\left(N\right) = N \int_0^{\infty} \mathrm{d}x\; \phi\left(x,n\right) q\left(x,n\right)^{N-1}.
\end{eqnarray}
Therefore, to compute the record rate, we need the more general survival and first-passage probabilities $q\left(x,n\right)$ and $\phi\left(x,n\right)$ for an arbitrary $x>0$.

Wergen et al.~\cite{Wergen2012} computed the asymptotic behavior of these quantities for $n\rightarrow\infty$ using a highly non-trivial theorem due to Ivanov \cite{Ivanov1996}, which is, in some sense, a very general form of Sparre Andersen theorem. To keep this essay readable, we will not describe the calculations of Wergen et al. and only summarize their results: The large $n$ behavior of $q\left(x,n\right)$ and $\phi\left(x,n\right)$ depends again on the L\'evy-index $\mu$ (see section \ref{rev:sec_biased_rw} and Eq. \ref{rev:f_k}) and one finds two universal regimes:

\paragraph{I --- The Brownian case with finite $\sigma^2$ ($\mu=2$): } Here, for $n\rightarrow\infty$, the first passage and survival probabilities approach the following forms:
\begin{eqnarray}
 \phi\left(x,n\right) \approx \frac{x}{\pi\sigma^2 n^{\frac{3}{2}}} e^{-\frac{x^2}{2\sigma^2 n}} \quad\quad\quad \textrm{and} \quad\quad\quad q\left(x,n\right) \approx \textrm{erf}\left(\frac{x}{\sqrt{2\sigma^2 n}}\right).
\end{eqnarray}
With these results one can compute the record rate directly. For a large number of random walks $N\gg1$ one finds:
\begin{eqnarray}
 P_n\left(N\right) \xrightarrow[n\rightarrow\infty]{N\rightarrow\infty} \sqrt{\frac{\ln N}{N}}.
\end{eqnarray}
Apparently, for $n,N\gg1$, the record rate of $N$ random walks is given by the record rate $P_n\left(N=1\right)$ of the single random walk times $\sqrt{\pi\ln N}$. Similarly, the mean record number of $N\gg1$ random walks approaches $\langle R_n\left(N\right)\rangle \approx \sqrt{4n\ln N} \approx \sqrt{\pi\ln N}\langle R_n\left(1\right)\rangle$.

\paragraph{II --- L\'evy flights with divergent $\sigma^2$ ($\mu<2$): } In this regime, it is not possible to compute the exact asymptotic behavior of $q\left(x,n\right)$ and $\phi\left(x,n\right)$. However, Wergen et al.~\cite{Wergen2012} could extract the scaling behavior of the product $\phi\left(x,n\right) q\left(x,n\right)^{N-1}$ in Eq. \ref{rev:P_n_N} and found that
\begin{eqnarray}
 P_n\left(N\right) \xrightarrow[n\rightarrow\infty]{N\rightarrow\infty} \frac{2}{\sqrt{\pi n}} \quad\quad\quad \textrm{and}\quad\quad\quad \langle R_n\left(N\right) \rangle \xrightarrow[n\rightarrow\infty]{N\rightarrow\infty} \frac{4\sqrt{n}}{\sqrt{\pi}}.
\end{eqnarray}
In this case, these quantities become completely independent from $N$ and are exactly twice as large as the corresponding results for $N=1$. The emergence of the prefactor $2$ and the complete independence of $N$ are very interesting findings, which are not entirely understood by now.

\bigskip
Wergen et al.~\cite{Wergen2012} also considered the distribution of the record number $P\left(R_n\left(N\right),n\right)$ in these two cases. As discussed before, because of the lacking renewal property, it is not possible to compute this distribution directly. However, in case I with $\mu=2$, it is possible to conjecture the asymptotic distribution $P\left(R_n\left(N\right),n\right)$ from the corresponding distribution of ensembles of random walks with a discrete jump distribution $f\left(x\right) = \frac{1}{2}\left(\delta\left(x+1\right)-\delta\left(x-1\right)\right)$. For such an ensemble of lattice random walks, the record number is simply given by the maximum $M_n\left(N\right)$ of the process. Since the maximum of a single lattice random walk has a finite variance, the maximum value of $N$ random walkers must be distributed according to a Gumbel distribution in the limit of large $n$ and $N$. In fact, one finds that the mean value of the maximum $\langle M_n\left(N\right)\rangle$ of $N$ random walkers converges to $\langle M_n\left(N\right)\rangle \approx \sqrt{2n\ln N}$. The mean record number of the $N$ random walkers with a continuous jump distribution only differs by a prefactor of $\sqrt{2}$. Using this analogy one can infer that the record number in the continuous case has the following Gumbel distribution:
\begin{eqnarray}
P\left(R_n\left(N\right)|n\right) \approx  e^{-z}e^{-e^{-z}}, \quad \textrm{with }z=\frac{R_n\left(N\right)-\sqrt{4n\ln N}}{\sqrt{n\ln N}}.
\end{eqnarray}
This conjecture was confirmed numerically in \cite{Wergen2012}.
For L\'evy flights with a heavy-tailed jump distribution ($\mu<2$) it was not possible to compute $P\left(R_n\left(N\right)|n\right)$ by similar means. However, performing numerical simulations, Wergen et al. \cite{Wergen2012} found that this distribution seems to be entirely independent from the L\'evy-index $\mu \in \left[0,2\right)$ and the number of walkers $N\gg1$. Because of this universality, it would be very interesting to find an analytical expression for this hitherto unknown distribution.

\subsection{Continuous-time random walks}

\begin{figure}[t]
\centerline{\includegraphics[width=0.75\textwidth]{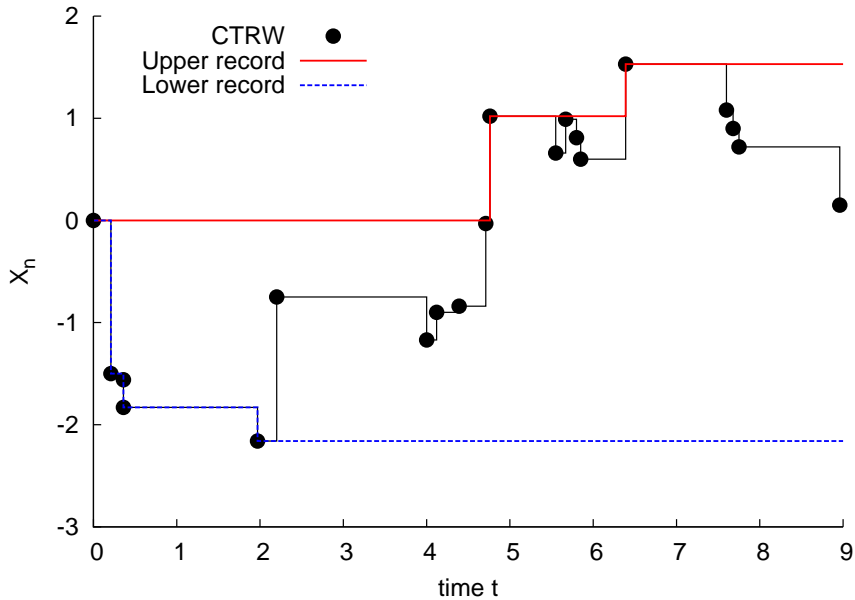}}
\caption{\label{rev:Fig_CTRW} Sketch of the record process of a continuous time random walk with random waiting times between the jumps (Here, we sampled the waiting times from an exponential distribution). The progression of the upper (lower) record is indicated by the red (blue dotted) line. }
\end{figure}

As another natural generalization of the symmetric DTRW studied by Majumdar and Ziff \cite{Majumdar2008} (section \ref{rev:symmetric_rw}), Sabhapandit \cite{Sabhapandit2011} studied continuous-time random walks (CTRW's). A CTRW is a process with entries $X\left(t_0\right),X\left(t_1\right),...,X\left(t_n\right)$ that are recorded at random times $t_0<t_1<...<t_n$ seperated by random waiting-times $\tau_i := t_{i}-t_{i-1}$ sampled from a (continuous) waiting time distribution $\rho\left(t\right)$ (see Fig. \ref{rev:Fig_CTRW}). In this context, a simple discrete-time random walk can be seen as a process with entries $X\left(0\right),X\left(1\right),...,X\left(n\right)$ at fixed times $t_0=0,t_1=1,...,t_n=n$ with a degenerate waiting-time distribution $\rho\left(t\right) = \delta\left(t-1\right)$. 

In the general case, the number of entries in a CTRW is apparently random, this makes it more complicated to consider the statistics of records. The number of records $R\left(t\right)$ at an arbitrary, continuous time $t$ is defined as \begin{eqnarray}
R\left(t\right) := \textrm{max}\{R_n|t_n \leq t\},
\end{eqnarray}

Sabhapandit discussed the record statistics of CTRW's in the limit of $t\rightarrow\infty$. He found, that the asymptotic behavior depends on the tail of the waiting-time distribution $\rho\left(\tau\right)$ and therefore introduced the Laplace transform $\tilde{\rho}\left(s\right)$ of $\rho\left(\tau\right)$:
\begin{eqnarray}
 \tilde{\rho}\left(s\right) := \int_0^{\infty} \mathrm{d}\tau\; \rho\left(\tau\right) e^{-s\tau}.
\end{eqnarray}
The behavior of the waiting-time distribution for large values of $t$ is encapsulated in the small $s\rightarrow0$ behavior of is Laplace transform. In general $\tilde{\rho}\left(s\right)$ can be expanded in powers of $s$ and, for small $s$, one finds $\tilde{\rho}\left(s\right) \approx 1 - \left(\tilde{\tau} s\right)^\alpha$ with parameters $\tilde{\tau}$ and $\alpha$ depending on the tail of the distribution $\rho\left(\tau\right)$. For waiting-time distributions with a finite mean value (in this case $\tilde{\tau}$) one finds $\alpha=1$, a heavy-tailed waiting-time distribution without a first moment yields an $\alpha$ between $0$ and $1$.

Sabhapandit showed that the first-passage probability of the CTRW can be obtained from the existing result for the discrete time random walk and computed the asymptotic record statistics for the two cases of $\alpha=1$ and $0<\alpha\leq1$.

In first case, for a finite mean waiting time $\tilde{\tau}$, the asymptotic record statistics of the CTRW is the same as in the time-discrete case. Here, for $t/\tilde{\tau}\rightarrow\infty$, the record number $R\left(t\right)$ is distributed according to the half-Gaussian distribution
\begin{eqnarray}
 P\left(R\left(t\right)|t\right) \approx \frac{1}{\sqrt{\pi}} \left(\frac{t}{\tilde{\tau}}\right)^{-\frac{1}{2}} \textrm{exp} \left(-\left(\frac{t}{\tilde{\tau}}\right)^{-1}\frac{R\left(t\right)^2}{4}\right), 
\end{eqnarray}
which, for $\tilde{\tau}=1$, is exactly the record number distribution found in the discrete case. Similarly, for $\alpha=1$, the mean record number and the record rate are given by
\begin{eqnarray}\label{rev:ctrw_alpha1}
 \langle R\left(t\right) \rangle \approx \frac{2}{\sqrt{\pi}} \left(\frac{t}{\tilde{\tau}}\right)^{\frac{1}{2}} \quad\quad\quad \textrm{and}\quad\quad\quad P\left(t\right) \approx \frac{1}{\sqrt{\pi}} \left(\frac{t}{\tilde{\tau}}\right)^{-\frac{1}{2}}.
\end{eqnarray}

In the case of a divergent mean waiting time with $\alpha<1$, the record number $R\left(t\right)$ approaches a different asymptotic distribution: 
\begin{eqnarray}
 P\left(R\left(t\right)|t\right) \approx \frac{2}{\alpha} \left(\frac{t}{\tilde{\tau}}\right)^{\frac{1}{2}} R\left(t\right)^{1-\frac{2}{\alpha}} L_{\frac{\alpha}{2}}\left(\left(\frac{t}{\tilde{\tau}}\right)^{-1} R\left(t\right)^{-\frac{2}{\alpha}}\right),
\end{eqnarray}
where $L_{\frac{\alpha}{2}}\left(z\right)$ is the pdf of a one-sided L\'evy-stable distribution, which, in general, can not be expressed analytically. The Laplace transform of $L_{\frac{\alpha}{2}}\left(z\right)$ is given by $\tilde{L}_{\frac{\alpha}{2}}\left(s\right) = e^{-s^{-\frac{\alpha}{2}}}$. For $\alpha<1$, the mean record number and the record rate grow more slowly with $n$. Sabhapandit \cite{Sabhapandit2011} found that, for an arbitrary $\alpha\in\left(0,1\right]$,
\begin{eqnarray}
 \langle R\left(t\right) \rangle \approx \frac{2}{\alpha \Gamma\left[\frac{\alpha}{2}\right]} \left(\frac{t}{\tilde{\tau}}\right)^{\frac{\alpha}{2}} \quad\quad\quad \textrm{and}\quad\quad\quad P\left(t\right) \approx \frac{1}{ \Gamma\left[\frac{\alpha}{2}\right]} \left(\frac{t}{\tilde{\tau}}\right)^{\frac{\alpha}{2}-1},
\end{eqnarray}
in good agreement with his findings for $\alpha=1$ in Eq. \ref{rev:ctrw_alpha1}.

\subsection{Records in higher-dimensional processes}

In 2011, Edery et al.~\cite{Edery2011}, considered DTRW's in two and three dimensions and discussed the record statistics of the distance of such a process from the origin. In the case of a one-dimensional random walk this distance at the time $n$ is just given by $|X_n|=\sqrt{X_n^2}$. Already in this case it was not yet possible to compute the exact record rate and the distribution of the record number analytically. 

Edery et al. were interested in DTRW's on an orthogonal lattice in two and three dimensions. At each time step, such a random walker jumps from one lattice site to an adjacent site in a random direction. They analyzed the number of records in the series of distances $|\vec{X_0}|,|\vec{X_1}|,...,|\vec{X_n}|$ from the origin using numerical simulations.

Edery et al. began with a discussion of a symmetric lattice random walk with a symmetric distribution of the jumps. In this case, they could demonstrate that the mean record number of this process has the same scaling behavior as in the case of the discrete-time random walk in one dimension. Without a bias the mean record number grows proportional to $\sqrt{n}$.

In the case of a biased lattice random walk, with a drift in an arbitrary direction, the asymptotic behavior changes. In all three considered dimensions, the asymptotic mean record numbers grows linearly in $n$.

\section{Applications}
\label{rev:applications}
     
\subsection{Climate records}
\label{rev:temperatures}

The most popular application of the theory of records in the last years was certainly the study of temperature records. The evident and most likely man-made increase of the global mean temperature over the last decades \cite{IPCC1} raised the question about the effects of this climatic change on the occurrence and the magnitude of extreme and record-breaking events \cite{Katz1984,Kerr1991,Easterling1997,Stott2004,Hoyt1981,Zorita2008,Rahmstorf2012,Bassett1992,Bassett1993}. While, it is intuitively clear to assume that a warming climate also leads to more heat and less cold records, the first systematic application of theoretical results from record statistics was presented by Benestad in 2003 and 2004 \cite{Benestad2003,Benestad2004}. He compared the record process of monthly temperature mean values from Scandinavian weathers stations with i.i.d.~RV's and found a small, but significant increase in the number of heat records. Interestingly, he also considered daily precipitation sums, where he could not determine any non-stationary behavior of the record rate \cite{Benestad2003,Benestad2006}.

In 2006, Redner and Peterson considered daily temperature measurements from a single weather station in Philadelphia. Even though their data set covered more than 100 years, they had difficulties to quantify the effect of global warming on the measurements from this station. Nevertheless, Redner and Peterson made important progress on the matter. In fact, they proposed a simple model of a Gaussian distribution with a linear drift to describe the record statistics of temperature measurements for individual calendar days. Within this model, a daily (mean, minimum or maximum) temperature $T_n$ in the $n$th year of an observation period is sampled from a Gaussian distribution with an increasing mean value $\mu_0 + ct$. Here, $c$ is the drift, which is basically the speed of warming. Then, the probability density of the daily temperatures should be of the form
\begin{eqnarray}
 f\left(T_n\right) = \frac{1}{\sqrt{2\pi\sigma}} e^{-\frac{\left(T_n - \mu_0 - ct\right)^2}{2\sigma^2}},
\end{eqnarray}
where $\sigma$ is the standard deviation that describes the fluctuations of the measurement around the moving mean value. Clearly, this is exactly the Linear Drift Model (LDM) we discussed in section \ref{rev:uncorr} for Gaussian RV's.

The work of Redner and Peterson motivated many others to study both the statistics of record breaking temperatures \cite{Meehl2009,Wergen2010,Newman2010,Anderson2011,Elguindi2012,Rahmstorf2011} and also the simple LDM \cite{Franke2010,Wergen2011,Franke2011b}. In 2009, Meehl et al.~\cite{Meehl2009} analyzed a large number of U.S.~weather stations with respect to the occurrence of heat and cold records and found a significant effect of global warming in the ratio of the two of them. In 2010, Wergen and Krug \cite{Wergen2010} confirmed these findings in an independent study of European station and re-analysis data \cite{USHCN,ECAD}. The work of Newman et al.~\cite{Newman2010}, Anderson and Kostinski \cite{Anderson2011}, Elguindi et al.~\cite{Elguindi2012} as well as Rahmstorf and Coumou \cite{Rahmstorf2011} lead to similar results. 

The main subject of these studies was a comparison between time series of temperature measurements for individual calendar days or months and uncorrelated RV's from the LDM or related, slightly more complicated models. In fact, it is by now well established that a Gaussian LDM, despite its simplicity, can describe the effect of global warming on the occurrence of daily temperature records relatively accurately. From 1960 to 2010, the global mean temperature increased in a roughly linear manner \cite{IPCC1,Stott2004}. This effect is also found in measurement from European and U.S. weather stations and re-analysis data. Over the same period, the standard deviation of the daily and monthly temperatures around their moving mean value remained more or less constant \cite{IPCC1,Stott2004,Wergen2010}. Since the magnitude of the warming in recent years is still much smaller than the average standard deviation of daily temperatures, one can compare the temperature measurements with a Gaussian LDM in the regime of small $c\ll \sigma/\sqrt{n}$ (see section \ref{rev:uncorr}). Here, with the findings of Franke et al.~\cite{Franke2011b}, the record rate $P_n\left(c\right)$ in a time series with a linear drift (warming) $c$ and standard deviation $\sigma$ is given by:
\begin{eqnarray}\label{rev:Gaussian_LDM_clim}
 P_n\left(c\right) \approx \frac{1}{n+1} + \frac{c}{\sigma} \frac{2\sqrt{\pi}}{e^2} \sqrt{\ln \left(\frac{n^2}{8\pi}\right)}.
\end{eqnarray}
Apparently, in this approximation, the important degree of freedom is the \textit{normalized drift} $\tilde{c} := \frac{c}{\sigma}$. For a fixed $n$, a large increase in the record rate can be caused by a large positive drift $c$, or a small standard deviation $\sigma$.

In most of the considered daily data sets, the normalized drift between 1960 and 2010 was of the order $\tilde{c} \approx 0.01\;\textrm{y}^{-1}$. For this value, Eq. \ref{rev:Gaussian_LDM_clim} predicts an increase in the record rate of about 27\% after 30 years and of more than 50\% after 50 years of warming. These predictions are in good agreement with the data found in the literature. For instance, Wergen et al.~\cite{Wergen2010} considered daily maximum temperatures measured at 202 European stations \cite{ECAD} between 1975 and 2005 and found an increase of around 40\% in the number of heat records along with a normalized drift of $\tilde{c}\approx 0.015\;\textrm{y}^{-1}$.

For monthly mean temperatures the normalized drift is usually larger, since the standard deviation of these averaged values is much smaller than in the case of daily measurements. Therefore, the rate of monthly upper records can be many times as high as expected in the case of a stationary climate \cite{Anderson2011,Wergen2012b}. Of course, for annual mean values this effect is even stronger and, over the last decade, the rate of new global mean temperature records was increased by around $2800\%$ \cite{Rahmstorf2012}. 

As shown by Wergen et al.~\cite{Wergen2010} and recently, in more detail, also by Elguindi et al.~\cite{Elguindi2012}, the normalized drift $\tilde{c}$ has strong regional variations. In Europe, $\tilde{c}$ seems to be generally larger than in the U.S., where $\tilde{c}$ for the daily data is usually much smaller than $0.01\;\textrm{y}^{-1}$ \cite{Wergen2010}. Due to the high heat capacity of water, the standard deviation of the daily measurements is much smaller near or over the oceans. Because of that, time series of these measurements can have a very large $\tilde{c}$ and therefore a very strong effect of global warming on the record rate. Stations far away from the sea, on the other hand, can have a very high standard deviation, which reduces the effect of the drift on the record rate. This explains, why the increase of the record rate in Europe was much stronger than in the U.S.~\cite{Wergen2010}: The temperature fluctuations at the U.S. stations, in particular of those in the middle of the continent, are much larger than at the European stations and therefore the normalized drift $\tilde{c}$ of the U.S. stations is much smaller. 

In a recent study, Wergen et al.~\cite{Wergen2012b} discuss the statistics of the values of record breaking temperatures. While the preceding literature focused only on the occurrence of temperature records, it is also possible to describe the effect of global warming on the values of these records using the LDM. However, here, a Gaussian LDM is appropriate only in the summer months, where the values of heat records are significantly increased by the warming. In the winter, especially in cold, sub-polar regions, the distribution of daily temperatures is highly asymmetric and favors extremely cold temperatures in comparison with the Gaussian case. Therefore, despite global warming, cold records are still much further away from the temperature mean value than heat records \cite{Wergen2012b}. In other words, despite of global warming, we can still expect extreme, record-breaking cold days in winter.

\subsection{Records in finance}
\label{rev:finance}

While, in the study of temperature records, the observational data was compared to uncorrelated random variables, the financial markets yield good examples of highly correlated processes. A simple way to model a stock, which is tradable at a stock market, is the Geometric Random Walk model (GRM), which was, in a slightly different form, already proposed by Le Bachelier in 1900 \cite{LeBachelier1900}. The GRM describes the logarithms of stocks prices with a simple biased random walk (as in section \ref{rev:sec_biased_rw}). In recent work Wergen et al.~\cite{Wergen2011b,Wergen2012a}, as well as Bogner \cite{Bogner2009}, discussed the record statistics of daily stock data from the U.S. Standard and Poors 500 (S\&P 500) stock index \cite{SPdata} in the context of this model.

They considered the logarithms of time series of daily stock prices $S_0,S_1,...,S_n$. Within the GRM, these logarithmic prices $\ln S_n$ should behave like a biased random walk with
\begin{eqnarray}
 \ln S_i = \ln S_{i-1} + \eta_i + c
\end{eqnarray}
with random jumps (daily returns) $\eta_i$ from a symmetric (return-) distribution $f\left(\eta\right)$ and a bias $c$. The bias represents a systematic, long-term growth in the system and leads, asymptotically, to an exponential growth of $S_n$. Since the logarithm is monotonic, a record in a series of stock prices $S_0,S_1,...,S_n$ is also a record in the series $\ln S_0,\ln S_1,...,\ln S_n$ and one can use the results for the record statistics of biased random walks to model the records of the stock prices.

As it turns out, the GRM is, to a certain degree, useful to predict the record number of daily stock prices from the S\&P 500. Wergen et al.~\cite{Wergen2011b} considered series of daily stock prices of 366 stocks from this index and analyzed the progression of the mean record number $\langle R_n\rangle$ of these stocks. 

To compare with the analytical findings for the biased random walk, they determined the drift $c$ and the standard deviation $\sigma$ of the jump distribution of the daily returns from the stock data. In the relevant regime of $c\sqrt{n} \ll \sigma$, they predicted a mean record number of (see section \ref{rev:sec_biased_rw})
\begin{eqnarray}\label{rev:R_n_finance}
 \langle R_n \rangle \approx \sqrt{\frac{4n}{\pi}} + \frac{1}{\sqrt{2}} \frac{c}{\sigma}.
\end{eqnarray}
Again, as in the case of temperature records in a warming climate, the relevant parameter that describes the effects of the bias is the normalized drift $\tilde{c} = \frac{c}{\sigma}$. For the S\&P 500 stock data one finds a value of $\tilde{c}$ between $0.018\;d^{-1}$ and $0.025\;d^{-1}$ depending on the length of the observation period.

Wergen et al.~\cite{Wergen2011,Wergen2012d} found that Eq. \ref{rev:R_n_finance} predicts the qualitative behavior of the mean record number of the stock prices to some accuracy. Even though it slightly overestimates both the number of upper and the number of lower records, the difference between the two is modeled correctly. 

In a similar study, Wergen et al.~\cite{Wergen2012} considered ensembles of multiple stocks from the S\&P 500 and compared them with their analytical results for multiple independent random walkers (section \ref{rev:sec_multiple_random_walks}). They rescaled and detrended the daily stock data to make them comparable with symmetric random walks with a jump distribution of standard deviation unity. Ensembles of $N$ of these rescaled stocks were then compared with the analytical result for the mean record number of the maximum of $N$ independent random walks (see section \ref{rev:sec_multiple_random_walks}):
\begin{eqnarray}
 \langle R_{n,N} \rangle \approx 2\sqrt{n\ln N}.
\end{eqnarray}
Interestingly, one finds that the maximum of $N$ detrended and rescaled stocks grows also proportional to $\sqrt{n\ln N}$, but has a different prefactor smaller than the one of the independent random walks. In \cite{Wergen2012} this was tentatively interpreted with a smaller, effective number of stocks that are stochastically independent in the context of record statistics. In view of the ongoing research on the important role of correlations between stocks in financial markets, it would be interesting to better understand the meaning of the effective number in the future.

\subsection{Physics and biology}
\label{rev:other_applications}

Interestingly, in physics and also in evolutionary biology, one finds several complex dynamical systems that behave like the record process of i.i.d.~RV's. In particular, some diffusive processes in random environments, like the random energy landscape by Derrida \cite{Derrida1981}, can be described using record statistics. These systems are usually stable on short time-scales, but run through intermittent events, so-called quakes, which bring them from one stable state to another. As it turns out these quakes can be modeled as record events in time series of i.i.d.~RV's.

An important feature of the record process of i.i.d.~RV's is that it can be described as a Poisson process in logarithmic time. According to Sibani and Littlewood \cite{Sibani1993} (see also the reviews by Jensen~\cite{Jensen2005} and Anderson et al.~\cite{Sibani2004}), the distribution of the logarithmic waiting times $\Delta_k := \ln t_k - \ln t_{k-1}$ between the $\left(k-1\right)$st and the $k$th record is given by the exponential distribution with the pdf $\rho\left(\Delta\right) = e^{-\Delta}$. With this one can show that the probability $P_k\left(t\right)$ of having $k\gg1$ records up to time $t\gg k$ is given by
\begin{eqnarray}\label{rev:log_Poisson}
 P_k\left(t\right) \approx \frac{1}{t}\frac{\left(\ln t\right)^{k-1}}{\left(k-1\right)!}
\end{eqnarray}
This kind of log-Poisson statistics is also found in various dynamical systems like, for instance, the Edwards Anderson spin-glass model \cite{Sibani2006,Sibani2007,Jensen2005}. The time-evolution of the (local) energy $E\left(t\right)$ of such a spin-glass, which relaxes towards a lower energy state after an initial quench, is characterized by a series of local energy minima $E_{\textrm{min}}\left(k\right)$ and maxima  $E_{\textrm{max}}\left(k\right)$. In order to get from one stable state with energy $E_{\textrm{min}}\left(k\right)$ to the next with energy $E_{\textrm{min}}\left(k+1\right)$, the system needs to overcome an energy barrier with $\Delta E_k = E_{\textrm{max}}\left(k\right) - E_{\textrm{min}}\left(k\right)$. Therefore, the noise driven system needs a fluctuation of the size $\Delta E_k$ to relax to the next stable state. Now, as shown by Sibani et al. \cite{Sibani2006,Sibani2007,Jensen2005}, these energy barriers $\Delta E_k$ are usually monotonically increasing in $k$ and one finds $\Delta E_k < \Delta E_{k+1}$. Because of that, the fluctuation necessary to overcome $\Delta E_{k+1}$ has to be (slightly) larger than the one needed for $\Delta E_k$. In other words, it requires a record-breaking event in the series of fluctuations for the system to relax further. Since these fluctuations are usually assumed to be i.i.d.~RV's, one can assume that the process of jumps (quakes) from one stable state to another has the same time-evolution as the record process of i.i.d.~RV's, which is describe by Eq.~\ref{rev:log_Poisson}.

A similar behavior is found in the so-called Restricted Occupancy Model, which was proposed in the context of the theory of type-II superconductors \cite{Sibani2004,Oliveira2005}. This model describes the gradual magnetization of a superconducting sample through an external magnetic field. In this context, the number of flux vortices inside a three dimensional model of the type-II superconductor increases step-wise and monotonically in time. The occurrence of the steps (quakes) in the vortex number also exhibits the same log-Poisson statistics as the record process of i.i.d.~RV's. 

As it turns out, the record process of i.i.d.~RV's can be used to describe various dynamical models related to the random energy model. In the context of evolutionary biology, several authors studied the connection between record statistics and adaptation on the fitness landscapes of genotypes. Such a landscape maps the fitness associated with a certain genotype to a (high-dimensional) cubic lattice similar to a lattice of spins. Kaufmann and Levin \cite{Kaufmann1998}, Sibani et al.~\cite{Sibani1998}, as well as Krug and Jain \cite{Krug2005} discussed mutations on random fitness landscapes with i.i.d. fitness values and compared the rate of their occurrence with the record rate of i.i.d.~RV's. The main idea is that, in order to survive and take over a population, a mutant with random fitness has to be fitter than all previous mutants in an evolutionary process. Therefore he must be a mutant with record-breaking fitness. 

A detailed discussion of the applications of record statistics in evolutionary biology can, for instance, be found in \cite{FrankeDiss}.

\subsection{Athletic records}
\label{rev:athletics}

Even though the occurrence of records in sports, such as in athletics or in swimming, receives an enormous amount of public attention, only very few have studied the statistical properties of these sport records so far. In the context of the ongoing controversy about the role of legal and illegal doping on the performance of athletes, the theory of records provides a method to distinguish between statistical fluctuations and real improvements. Because of the universal features of the record statistics of i.i.d.~RV's, one can analyze the occurrence of records in time series of sports results without detailed knowledge about the underlying distribution from which the results are sampled. In principle, if the number of records, in a series of sports results, is significantly larger (or smaller) than in an i.i.d.~series of comparable length, this can not be in agreement with a constant performance level of the athletes. However, when analyzing historical data, it is hard to determine the total number of attempts in a certain event and it is therefore difficult to determine the record rate $P_n$. Usually only a small number of very good performances is recorded on the leaderboard and one can only analyze the statistics of their values.

By now, the only systematic analysis of athletic records was published by Gembris et al.~\cite{Gembris2002,Gembris2007}. They considered the evolution of the record values of several track and field events and compared them to theoretical results for the maximal values of Gaussian i.i.d.~RV's. They estimated the mean value and the standard deviation of the athletic performances for the time series of individual events. Then they compared the record events in the athletic data with series of Gaussian RV's with the same parameters. A comparison of the record values allows to identify events, where the athletes improved significantly over the duration of the time series. It turns out that only in some events the record values actually improve faster than expected on the basis of constant athletic capabilities. In 50\% to 80\% of all considered track and field events, Gembris et al.~\cite{Gembris2002,Gembris2007} could not disprove their null hypothesis of a stationary distribution of athletic performances. Interestingly, in the cases where they could detect a systematic time-dependence, the increase in the performance seemed to be far from linear in time. If the performances of athletes would improve due to better training, nutrition, or just a growing population, one would expect a continuous effect on the record rate. Instead, the progressions of some athletic record values, especially in long distance running and in throwing, are characterized by large jumps, which are probably best explained by instantaneous effects like the introduction, or the prohibition of certain drugs. In fact, while the record values of several long distance running events, like 5000m or 10000m improved drastically after the introduction of blood doping with erythropoietin, almost no records were set in the throwing events since anabolic drugs became detectable in urine samples \cite{Gembris2002,Gembris2007}.

Another interesting problem in this context, is the question about an absolute limit to world records in sports, like, for instance, a hypothetical speed that can not be exceeded by humans, which would lead to a lower boundary for the possible outcome of a 100m dash race. Up to now, several different methods were applied to find such a boundary, but the issue is still controversial \cite{Deakin1967,Smith1988,Nevill2005,Lippi2008}. It might be possible to answer this question using extreme-value or record statistics, since these exhibit different universal properties for distributions with a bounded and an infinite support. A first step towards this goal was made by Einmahl and Magnus in 2008 \cite{Einmahl2008}. They estimated the tail behavior of the distributions of performances in track and field events by comparing them to a generalized Pareto distribution. In most considered cases, they could find an absolute limit to the world records. It would interesting to confirm and improve their findings in future studies.

\section{Summary and outlook}
\label{rev:summary}

In this review, we tried to summarize numerous interesting and non-trivial results on the statistics of records, which were discovered in the last couple of decades. Especially in the last 10 years, the study of record-breaking events has become a broad and diverse field of research. Additionally, the occurrence and the properties of records were analyzed and discussed in a vast number observational data sets. Researchers have understood that records are often more than just interesting to the observer: One can learn a lot about the properties of a complex dynamical system by considering the record events it generates.

In this context, there are many open problems which might suit as subjects for future research. The research on the record statistics of uncorrelated random variables with time-dependent distributions has just began and only the very simple cases of a constant linear drift and an increasing standard deviation have been understood to some degree. Of course, one can ask, how the record rate and also the full distribution of the record number is affected by a more complicated time dependence of the underlying distribution. For the Linear Drift Model and the Increasing Variance Model discussed in this review, it would also be interesting to compute the mean values of records that have a certain record number, or occur at a certain time. 

Furthermore, the record statistics of correlated random variables are only understood for a few special cases. Especially in the context of possible applications in finance, it would be very interesting to calculate the record rate of more complicated non-Markovian processes. Some interesting candidates for future research are branching random walks, the absolute value of a random walk and the Ornstein-Uhlenbeck process, which is particularly important for the modeling of financial data \cite{Barndorff2001}. In general, it would be desirable to better understand the effect of long-term, or power-law type correlations on the record statistics of stochastic processes.

With respect to the various applications of the theory one easily finds numerous interesting open question. For instance in climatology, it is hardly understood how specific weather conditions affect the occurrence of records. Here, it is also still unclear if one can find a significant effect of climatic change on the record statistics of precipitation events or, for instance, also record-breaking storms. In finance, our understanding of the statistics of record-breaking stock prices does still not explain some interesting deviations from the classical model of a geometric random walk. It is a challenging problem to find a more accurate description for this record process.

\subsubsection*{Acknowledgements}

The author is grateful to Joachim Krug and Jasper Franke for many helpful discussions and support on the subject of this review and Ivan Georg Szendro Ter\'an for his input towards the completion of this work. 

\section*{References}
\bibliographystyle{iopart-num}
\bibliography{lit_review}{}

\end{document}